\documentclass[fleqn,usenatbib]{mnras}

\usepackage[T1]{fontenc}
\usepackage{ae,aecompl}
\usepackage{threeparttable}
\usepackage{graphicx}	% Including figure files
\usepackage{amsmath}	% Advanced maths commands
\usepackage{amssymb}	% Extra maths symbols

\usepackage{newtxtext,newtxmath}
\title[\mbox{H\,{\sc i}} absorption towards PKS~0409$-$75]{\mbox{H\,{\sc i}} absorption at $z\sim 0.7$ against the lobe of the powerful radio galaxy PKS\,0409$-$75}

\author[E. K. Mahony et al.]{Elizabeth K. Mahony$^{1,2}$\thanks{E-mail: elizabeth.mahony@csiro.au}, 
James R. Allison$^{3,2}$,
Elaine M. Sadler$^{1,2,4}$,
Sara L. Ellison$^{5}$, \newauthor
Sui Ann Mao$^{6}$, 
Raffaella Morganti$^{7,8}$, 
Vanessa A. Moss$^{1,4}$,
Amit Seta$^{9}$, 
Clive N. Tadhunter$^{10}$,\newauthor
Simon Weng$^{1,4}$,
Matthew T. Whiting$^{1}$,
Hyein Yoon$^{2,4}$, 
Martin Bell$^{11}$,
John D. Bunton$^{1}$,\newauthor
Lisa Harvey-Smith$^{12,13}$,
Amy Kimball$^{14}$,
B\"arbel S. Koribalski$^{1,13}$
and Max A. Voronkov$^{1}$
\\
$^{1}$Australia Telescope National Facility, CSIRO Space and Astronomy, PO Box 76, Epping, NSW 1710, Australia. \\
$^{2}$ARC Centre of Excellence for All-Sky Astrophysics in 3 Dimensions (ASTRO 3D) \\
$^{3}$Sub-Dept. of Astrophysics, Department of Physics, University of Oxford, Denys Wilkinson Building, Keble Rd., Oxford, OX1 3RH, UK \\
$^{4}$Sydney Institute for Astronomy, School of Physics A28, The University of Sydney, NSW 2006, Australia. \\
$^{5}$Department of Physics \& Astronomy, University of Victoria, Finnerty Road, Victoria, British Columbia, V8P 1A1, Canada\\
$^{6}$Max Planck Institute for Radio Astronomy, Auf dem H\"{u}gel 69, 53121 Bonn, Germany \\
$^{7}$ASTRON, the Netherlands Institute for Radio Astronomy, Postbus 2, 7990 AA, Dwingeloo, The Netherlands.\\
$^{8}$Kapteyn Astronomical Institute, University of Groningen, Postbus 800, 9700 AV Groningen, The Netherlands.\\
$^{9}$Research School of Astronomy \& Astrophysics, The Australian National University, Canberra ACT 2611, Australia\\
$^{10}$Department of Physics \& Astronomy, University of Sheffield, Sheffield S3 7RH.\\  
$^{11}$University of Technology Sydney, 15 Broadway, Ultimo NSW 2007, Australia. \\
$^{12}$School of Physics, University of New South Wales, Sydney, NSW 2052, Australia. \\
$^{13}$School of Computer, Data and Mathematical Sciences, Western Sydney University, Locked Bag 1797, Penrith, NSW, 2751, Australia. \\
$^{14}$National Radio Astronomy Observatory, 1003 Lopezville Rd., Socorro, NM, 87801, USA. 
}

\date{Accepted XXX. Received YYY; in original form ZZZ}

\pubyear{2021}

\begin{document}
\label{firstpage}
\pagerange{\pageref{firstpage}--\pageref{lastpage}}
\maketitle

\begin{abstract}

We present results from a search for the \mbox{H\,{\sc i}} 21-cm  line in absorption towards 16 bright radio sources with the 6-antenna commissioning array of the Australian Square Kilometre Array Pathfinder (ASKAP). Our targets were selected from the 2-Jy sample, a flux-limited survey of the southern radio sky with extensive multi-wavelength follow-up. Two sources were detected in \mbox{H\,{\sc i}} absorption including a new detection towards the bright FRII radio galaxy PKS\,0409$-$75 at a redshift of $z=0.674$. The \mbox{H\,{\sc i}} absorption line is blueshifted by $\sim$3300 km\,s$^{-1}$ compared to the optical redshift of the host galaxy of PKS\,0409$-$75 at $z=0.693$. Deep optical imaging and spectroscopic follow-up with the GMOS instrument on the Gemini-South telescope reveal that the \mbox{H\,{\sc i}} absorption is associated with a galaxy in front of the southern radio lobe with a stellar mass of $3.2 - 6.8 \times 10^{11}M_\odot$, a star-formation rate of $\sim 1.24 M_\odot$\,yr$^{-1}$ and an estimated \mbox{H\,{\sc i}} column density of $2.16\times10^{21}$\,cm$^{-2}$, assuming a spin temperature of $T_{\rm spin}=500$\,K and source covering factor of $C_{\rm f}=0.3$. Using polarisation measurements of PKS\,0409$-$75 from the literature we estimate the magnetic field of the absorbing galaxy to be $\sim 14.5\mu$G, consistent with field strengths observed in nearby spiral galaxies, but larger than expected for an elliptical galaxy. Results from this pilot study can inform future surveys as new wide-field telescopes allow us to search for 21-cm \mbox{H\,{\sc i}} absorption towards all bright radio sources as opposed to smaller targeted samples.

\end{abstract}


%%%%%%%%%%%%%%%%%%%%%%%%%%%%%%%%%%%%%%%%%%%%%%%%%%

%%%%%%%%%%%%%%%%% BODY OF PAPER %%%%%%%%%%%%%%%%%%

\section{Introduction}

It is well established that the brightest extragalactic radio sources are produced by synchrotron-emitting radio jets launched from a central active galactic nucleus (AGN; \citealt{Condon1989, nvss, Mauch2007, Condon2019}). The precise conditions required to trigger and form these jets remains an active area of research, but the presence of cold gas as a source of fuel for the central AGN is a critical component. As such, studying the distribution and kinematics of cold gas at the cores of active galaxies is crucial to understanding the fuelling and feedback processes that shape their evolution. Observing the dynamics and kinematics of the gas can provide some of the most direct evidence for fuelling \citep{Maccagni2014, Tremblay2016, Chandola2020} and feedback processes \citep{Morganti2005, Dasyra2012, Morganti2013, Tadhunter2014, Oosterloo2019, HerraraCamus2020, Schulz2021} in AGN. 

Searching for \mbox{H\,{\sc i}} 21-cm absorption against bright radio sources is a highly effective tool in probing the dynamics of the cool, neutral gas, particularly beyond the nearby Universe ($z > 0.2$) where emission line studies require prohibitively long integration times (see \citealt{Morganti2015, Morganti2018} and references therein). In addition to being detectable out to large redshift, 21-cm absorption line studies have the advantage of directly tracing the gas at the position of the radio AGN, are more sensitive to the colder gas, and have the potential to probe gas at higher spatial resolution (e.g. \citealt{Beswick2002, Beswick2004, Morganti2013, Schulz2018}).

\citet{VanGorkom1989} carried out one of the first systematic searches for \mbox{H\,{\sc i}} absorption in radio-loud AGN with the Very Large Array (VLA), detecting \mbox{H\,{\sc i}} in 13\% of the 29 radio galaxies observed. Since then, there have been numerous studies searching for \mbox{H\,{\sc i}} absorption using different source selections \citep{Morganti2001, Gupta2006, Curran2008, Gereb2015, Allison2015, Maccagni2017, Murthy2021}. These searches typically detect the presence of neutral gas in approximately 30\% of sources \citep{Gereb2015,Maccagni2017}, although this varies significantly according to the sample selection. Numerous studies have reported higher detection rates in compact radio sources ranging from 32-57\% \citep{Vermeulen2003, Pihlstrom2003, Gupta2006, Chandola2013, Maccagni2017, Murthy2021} while a study of galaxy mergers found \mbox{H\,{\sc i}} absorption in 84\% of the sample \citep{Dutta2018,Dutta2019}. By comparison, \citet{Maccagni2017} report a detection rate of $\sim$15\% for extended radio sources and \citet{Murthy2021} did not detect any \mbox{H\,{\sc i}} 21-cm absorption lines towards the 14 resolved radio sources in their sample.

The majority of these studies have been carried out at low redshifts largely due to the available frequency coverage of most interferometers, such as the Westerbork Synthesis Radio Telescope (WSRT; \citealt{Gereb2015, Maccagni2017, Vermeulen2003}), the VLA  \citep{VanGorkom1989, Morganti2005, Murthy2021} and the Australia Telescope Compact Array (ATCA; \citealt{Allison2012a, Glowacki2017}). In addition, the relatively narrow bandwidths meant that prior knowledge of the redshift of the radio source was needed to ensure the \mbox{H\,{\sc i}} 21-cm line was observable in the frequency range available. This was particularly true for observations carried out prior to significant upgrades of the ATCA and VLA which increased the available bandwidth from $2\times128$\,MHz to $2\times2048$\,MHz frequency windows for ATCA \citep{cabb} and from 100\,MHz to 2--8\,GHz (depending on the spectral resolution and frequency coverage) for the VLA \citep{evla}. This requirement of prior knowledge of the spectroscopic redshift introduces various selection biases since the optical magnitude of the galaxy needs to be sufficiently bright to obtain a redshift measurement. An increasing number of high-redshift searches ($z >1 $) have been performed using the Giant Metrewave Radio Telescope (GMRT; \citealt{Curran2008, Curran2012, Aditya2019, Aditya2021}), significantly enhanced thanks to the recent upgrade which increased the instantaneous bandwidth from 32\,MHz to 400\,MHz and increasing the available frequency coverage from 50-1500\,MHz \citep{Aditya2016, Aditya2019, Dutta2020}. 

The advent of new facilities with wider bandwidths, including the Australian Square Kilometre Array Pathfinder (ASKAP; \citealt{Hotan2021}), MeerKAT \citep{Jonas2016}, Westerbork Aperture Tile in Focus (AperTIF; \citealt{apertif}) and the upgraded GMRT \citep{Gupta2017}, have opened up a much larger frequency window to search for \mbox{H\,{\sc i}} in absorption. Importantly, the large instantaneous bandwidth ($\sim300$\,MHz for ASKAP and Apertif and $\sim400$\,MHz for uGMRT and MeerKAT), and relatively Radio Frequency Interference (RFI)-free spectrum allow us to search for absorption lines without prior knowledge of the redshift. This has been demonstrated by the recent detection of several \mbox{H\,{\sc i}} absorption systems where the redshift of the background radio source was not known before the observations \citep{Allison2015, Moss2017, Glowacki2019, Allison2020, Chowdhury2020, Sadler2020}. 

This paper presents a search for \mbox{H\,{\sc i}} absorption towards 16 powerful radio galaxies with the Boolardy Engineering Telescope Array (BETA); the 6-element commissioning array of ASKAP \citep{Hotan2014, Mcconnell2016}. The sources were selected from the 2-Jy sample \citep{WallPeacock1985, Tadhunter1993} to ensure there was sufficient radio continuum to be able to detect \mbox{H\,{\sc i}} absorption with the 6 antennas of ASKAP-BETA. Section 2 describes the sample selection, observations and data reduction of the ASKAP-BETA observations and Section 3 presents the results of the ASKAP-BETA search for \mbox{H\,{\sc i}} absorption.  Section 4 describes follow-up optical observations carried out with the Gemini-South telescope. We discuss the properties of the \mbox{H\,{\sc i}} absorption detection and implications for similar detections in upcoming large absorption line surveys in Section 5 before concluding in Section 6. 

\section{Radio Observations}

\subsection{Sample selection}

A sample of 16 bright radio galaxies was selected from the 2-Jy sample \citep{WallPeacock1985}; a complete, flux-density limited sample of sources brighter than 2\,Jy at 2.7\,GHz. An extensive campaign of multi-wavelength follow-up has been carried out for 2-Jy sources at declinations $\delta < +10^{\circ}$, steep spectral indices ($\alpha_{2.7}^{4.8}<-0.5$)\footnote{where $S_{\nu}\propto\nu^{\alpha}$ and the sub/super-scripts refer to the frequency range in GHz between which the spectral index was measured} and redshifts between $0.05<z<0.7$ \citep{Tadhunter1993}. This includes high-resolution radio imaging \citep{Morganti1999}, deep optical imaging and spectroscopy \citep{RamosAlmeida2011}, mid and far-IR imaging \citep{Dicken2008, Dicken2012} and X-ray observations \citep{Mingo2014}. The completeness of the multiwavelength follow-up make it an ideal sample to search for \mbox{H\,{\sc i}} absorption as it is a radio-selected sample, yet has the accompanying multi-wavelength data needed to provide redshifts and luminosities without introducing additional selection effects based on the optical properties (e.g. such as those introduced by selecting sources with existing spectroscopic data from large optical surveys which can add an additional optical flux-limit). Only sources with redshifts $z>0.2$ were included to ensure the \mbox{H\,{\sc i}} line would fall in the frequency range observed which can trace the \mbox{H\,{\sc i}} line from $0.2<z<1$. 

A summary of the properties of these 16 radio galaxies is given in Table \ref{sampletab}. All sources have been classified according to their radio morphology (Fanaroff-Riley Type I/II (FRI/FRII) or Compact Steep Spectrum (CSS) objects) and their optical spectral properties (Narrow or Broad Line Radio Galaxy (NLRG/BLRG, QSO) given in \citet{Tadhunter1993, Morganti1993}. We also list the measured flux densities at 4.8\,GHz (both the core and total flux measurements as given in \citet{Morganti1993}) and at 850\,MHz from the Rapid ASKAP Continuum Survey (RACS; \citealt{McConnell2020, Hale2021}).

\begin{table*}
\caption{Properties of the 2-Jy sources that were searched for \mbox{H\,{\sc i}} absorption with ASKAP-BETA. The radio morphology class refers to either a Fanaroff-Riley Type I/II radio galaxy or a Compact Steep Spectrum Source (CSS) and are taken from \citet{Morganti1993}, as are the total and core flux densities measured at 4.8\,GHz. The optical classification is determined from the spectroscopic properties and defined as either a Quasi-Stellar Object (QSO), Narrow-line or Broad-line Radio Galaxy (NLRG/BLRG) as designated in \citet{Tadhunter1993}. The final column lists the 850\,MHz flux density from the Rapid ASKAP Continuum Survey (RACS). All flux densities are listed in Jy. \label{sampletab}}
\begin{threeparttable}
\begin{tabular}{lclcccrcr}
\hline
Source & RA (J2000) & DEC (J2000) & $z$ & Radio cl. & Opt. cl. & $S_{\rm{tot,4.8\,GHz}}$ & $S_{\rm{core,4.8\,GHz}}$ & $S_{\rm{850\,MHz}}$ \\%& $R_{\rm {2.3 GHz}}$ \\
\hline
PKS\,0023$-$26 & 00$^{\rm h}$25$^{\rm m}$49$\fs$16 & $-$26$^\circ$02$\arcmin$12$\farcs$6 & 0.322 & CSS & NLRG & 3.41 & unresolved & 11.6 \\%& --- \\ 
PKS\,0035$-$02 & 00$^{\rm h}$38$^{\rm m}$20$\fs$52 & $-$02$^\circ$07$\arcmin$40$\farcs$7 & 0.220 & (FRII)$^*$ & BLRG & 2.64 & 0.662 & 9.1 \\%& 0.1133 \\
PKS\,0039$-$44 & 00$^{\rm h}$42$^{\rm m}$08$\fs$98 & $-$44$^\circ$14$\arcmin$00$\farcs$5 & 0.346 & FRII & NLRG & 1.17$^\dagger$ & no core & 5.7 \\%& ---\\
PKS\,0105$-$16 & 01$^{\rm h}$08$^{\rm m}$16$\fs$90 & $-$16$^\circ$04$\arcmin$20$\farcs$6 & 0.400 & FRII & NLRG & 1.17 & $<$0.002 & 6.3 \\%& $<$0.005\\ 
PKS\,0117$-$15 & 01$^{\rm h}$20$^{\rm m}$27$\fs$10 & $-$15$^\circ$20$\arcmin$16$\farcs$6 & 0.565 & FRII & NLRG & 1.6 & no core & 7.1 \\%& $<$0.004\\
PKS\,0235$-$19 & 02$^{\rm h}$37$^{\rm m}$43$\fs$45 & $-$19$^\circ$32$\arcmin$33$\farcs$3 & 0.620 & FRII & BLRG & 1.44 & $<$0.0002 & 6.8 \\%& $<$0.0002\\
PKS\,0252$-$71 & 02$^{\rm h}$52$^{\rm m}$46$\fs$16 & $-$71$^\circ$04$\arcmin$35$\farcs$3 & 0.563 & CSS & NLRG & 1.58 & unresolved & 8.8 \\%& ---\\
PKS\,0409$-$75 & 04$^{\rm h}$08$^{\rm m}$48$\fs$49 & $-$75$^\circ$07$\arcmin$19$\farcs$3 & 0.693 & FRII & NLRG & 4.25 & no core & 19.6 \\%& --- \\
PKS\,1136$-$13 & 11$^{\rm h}$39$^{\rm m}$10$\fs$70 & $-$13$^\circ$50$\arcmin$43$\farcs$6 & 0.554 & FRII & QSO & 1.9 & no core & 6.2 \\%& 0.1551\\
PKS\,1151$-$34 & 11$^{\rm h}$54$^{\rm m}$21$\fs$79 & $-$35$^\circ$05$\arcmin$29$\farcs$1 & 0.258 & CSS & QSO & 2.78 & unresolved & 7.6 \\%& --- \\
PKS\,1306$-$09 & 13$^{\rm h}$08$^{\rm m}$39$\fs$12 & $-$09$^\circ$50$\arcmin$32$\farcs$5 & 0.464 & CSS & NLRG & 1.9 & unresolved & 5.6 \\%& --- \\
PKS\,1547$-$79 & 15$^{\rm h}$55$^{\rm m}$21$\fs$65 & $-$79$^\circ$40$\arcmin$36$\farcs$3 & 0.483 & FRII & BLRG & 1.38 & $<$0.003 & 6.1$^\ddag$ \\%& 0.0071 \\
PKS\,1602$+$01 & 16$^{\rm h}$04$^{\rm m}$45$\fs$32 & $+$01$^\circ$17$\arcmin$51$\farcs$0 & 0.462 & FRII & BLRG & 1.13 & 0.065 & 6.8 \\%& 0.0023 \\
PKS\,1938$-$15 & 19$^{\rm h}$41$^{\rm m}$15$\fs$07 & $-$15$^\circ$24$\arcmin$31$\farcs$3 & 0.452 & FRII & BLRG & 2.34 & no core & 9.2\\%& ---\\
PKS\,2135$-$14 & 21$^{\rm h}$37$^{\rm m}$45$\fs$17 & $-$14$^\circ$32$\arcmin$55$\farcs$8 & 0.200 & FRII & QSO & 1.38 & 0.121 & 5.8\\%& 0.0633 \\
PKS\,2135$-$20 & 21$^{\rm h}$37$^{\rm m}$50$\fs$01 & $-$20$^\circ$42$\arcmin$31$\farcs$6 & 0.635 & CSS & BLRG & 1.53 & unresolved & 5.2\\%& --- 
\hline
\end{tabular}
\begin{tablenotes}
\item$^*$ Brackets indicate an uncertain radio morphology for this object.
\item$^\dagger$ Flux densities published in \citet{Morganti1999} for this object.
\item$^\ddag$ Flux density for this source was obtained from the initial RACS catalogue available from the CSIRO ASKAP Science Data Archive (CASDA)\footnote{https://research.csiro.au/casda/}.
\end{tablenotes}
\end{threeparttable}
\end{table*}

\subsection{ASKAP-BETA}

Observations were carried out using ASKAP-BETA over the period from November 2014 - February 2016. This 6-element interferometer had baselines ranging from 37 to 916m, leading to a synthesised beam of approximately 1.5\,arcmin at 850\,MHz when using natural weighting. 

The majority of sources were observed over the frequency range 711.5 -- 1015.5 MHz corresponding to \mbox{H\,{\sc i}} redshifts in the range  $0.4<z<1.0$. Five targets were observed at higher frequencies (967.5 -- 1271.5 MHz) which covers the redshift range $0.2<z<0.4$, however, this frequency range is more strongly affected by RFI from satellites and aircraft, with approximately 30-50\% of the data being flagged. Each target was placed at the phase centre of the central beam, with the remaining 8 beams arranged in a diamond footprint (see \citealt{Allison2015} for more details). The primary flux calibrator, PKS\,B1934$-$638, was observed for short scans of 5--15 min in each beam at the beginning of each observing run. A summary of the ASKAP-BETA observations is given in Table \ref{obstab}. 

The data were reduced using a custom-built {\sc miriad} pipeline as described in detail by \citet{Allison2015}. In summary, the data were separated into 64 subbands corresponding to the 4-5\,MHz beam-forming frequency intervals used by ASKAP-BETA and each subband then reduced independently using the standard flagging, calibration and imaging {\sc miriad} tasks. Processing each subband separately provided the advantage of allowing parallelisation of the data reduction, but also corrected for the discrete jumps in the gain solutions at the edge of the beam-forming intervals. 

The continuum subtraction was performed in two steps: first, by subtracting the {\sc clean} components generated from a continuum image of each subband from the visibilities using {\sc uvmodel}, and then by fitting a second-order polynomial using {\sc uvlin} to subtract any residual flux from continuum sources that were not sufficiently removed from the visibilities. Finally, a spectrum was extracted at the peak flux density of the target source (which were all unresolved by ASKAP-BETA) at the full spectral resolution of 18.5\,kHz which corresponds to a rest-frame velocity resolution of 5.5-7.8\,km\,s$^{-1}$. For consistency, the data were re-reduced after the completion of all observations to ensure the same version of the pipeline was used for all targets.

After reducing each observation independently, the extracted spectrum for each target was averaged and weighted by the inverse-variance measured across the channel images. For more details on the data processing we refer the reader to \citet{Allison2015}.

\section{Results}

\begin{figure*}
\vspace{-0.3cm}
\begin{minipage}{0.48\linewidth}
\centering{\includegraphics[width=\linewidth]{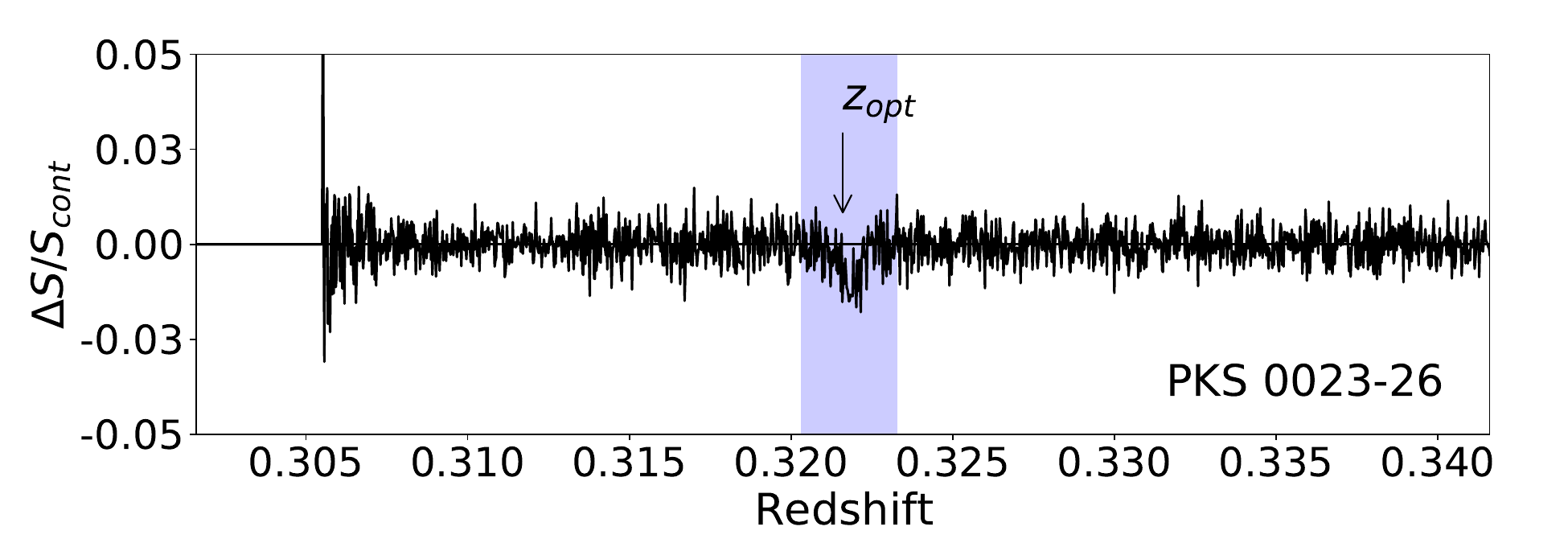}}
\end{minipage}
\vspace{-0.13cm}
\begin{minipage}{0.48\linewidth}
\centering{\includegraphics[width=\linewidth]{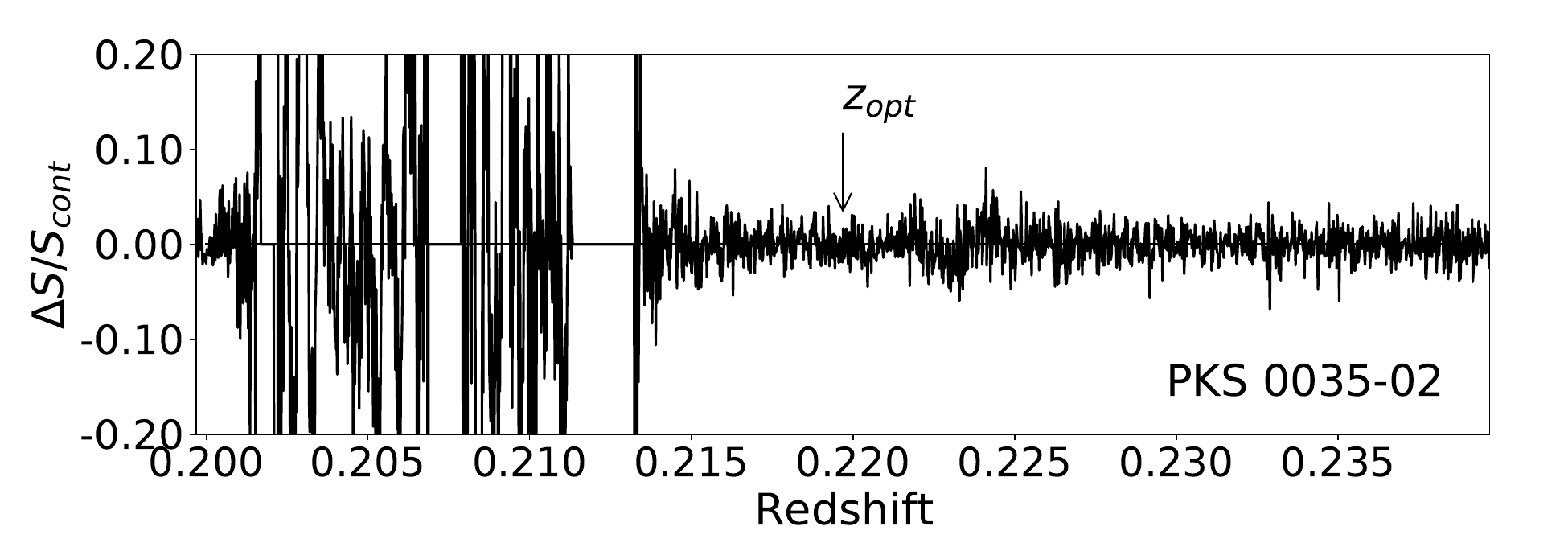}}
\end{minipage}
\vspace{-0.13cm}
\begin{minipage}{0.48\linewidth}
\centering{\includegraphics[width=\linewidth]{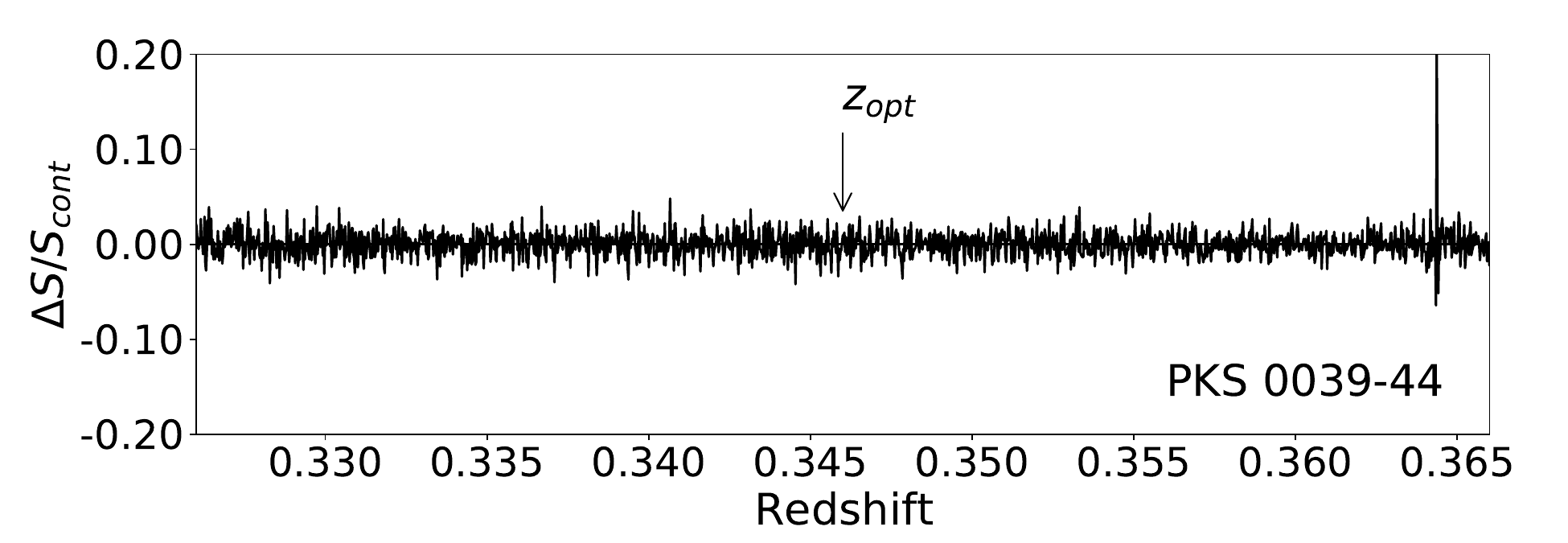}}
\end{minipage}
\begin{minipage}{0.48\linewidth}
\centering{\includegraphics[width=\linewidth]{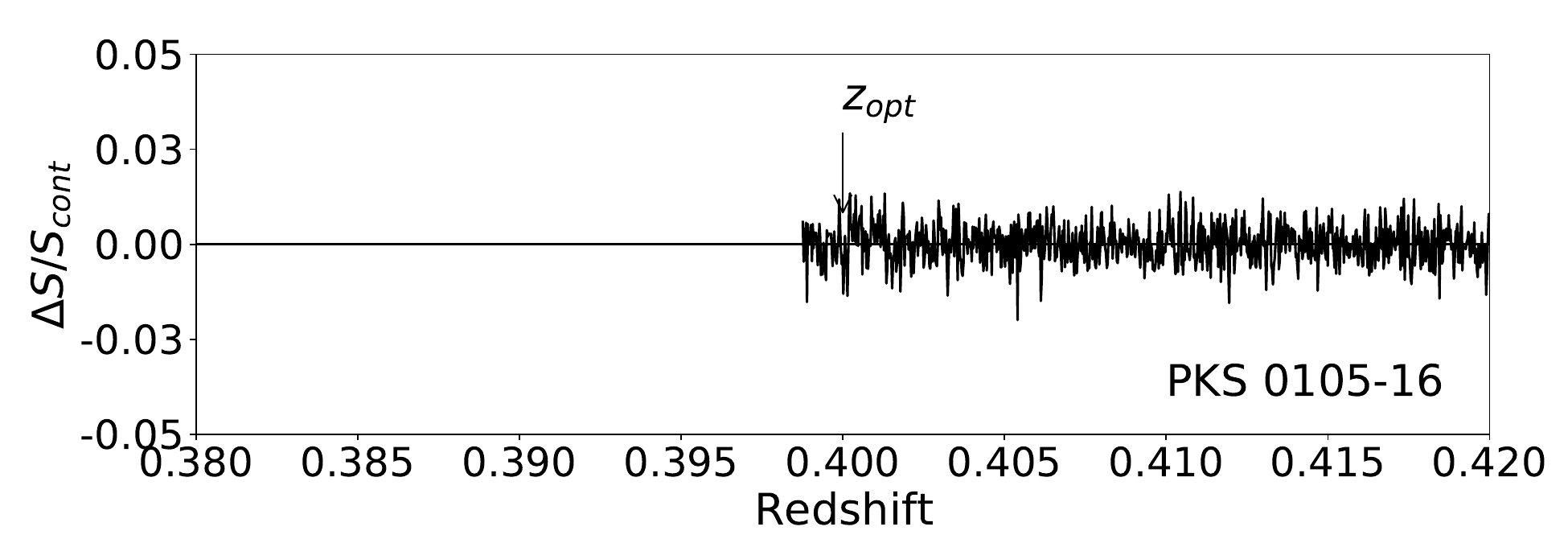}}
\end{minipage}
\vspace{-0.13cm}
\begin{minipage}{0.48\linewidth}
\centering{\includegraphics[width=\linewidth]{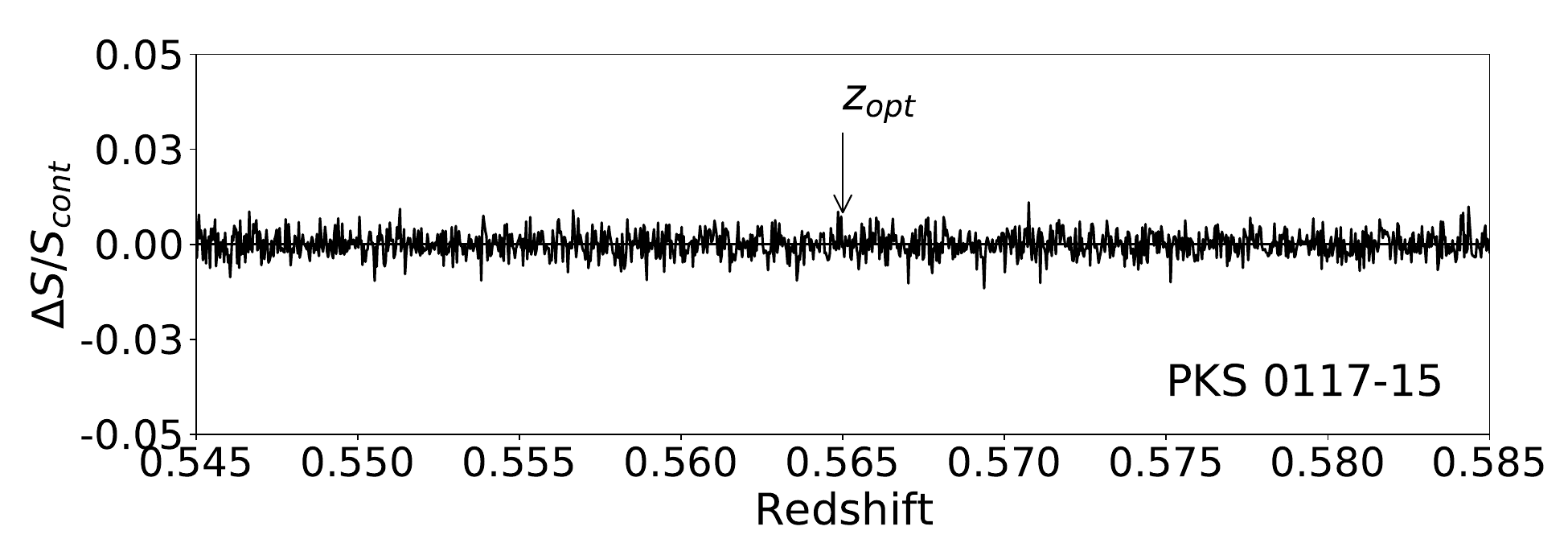}}
\end{minipage}
\begin{minipage}{0.48\linewidth}
\centering{\includegraphics[width=\linewidth]{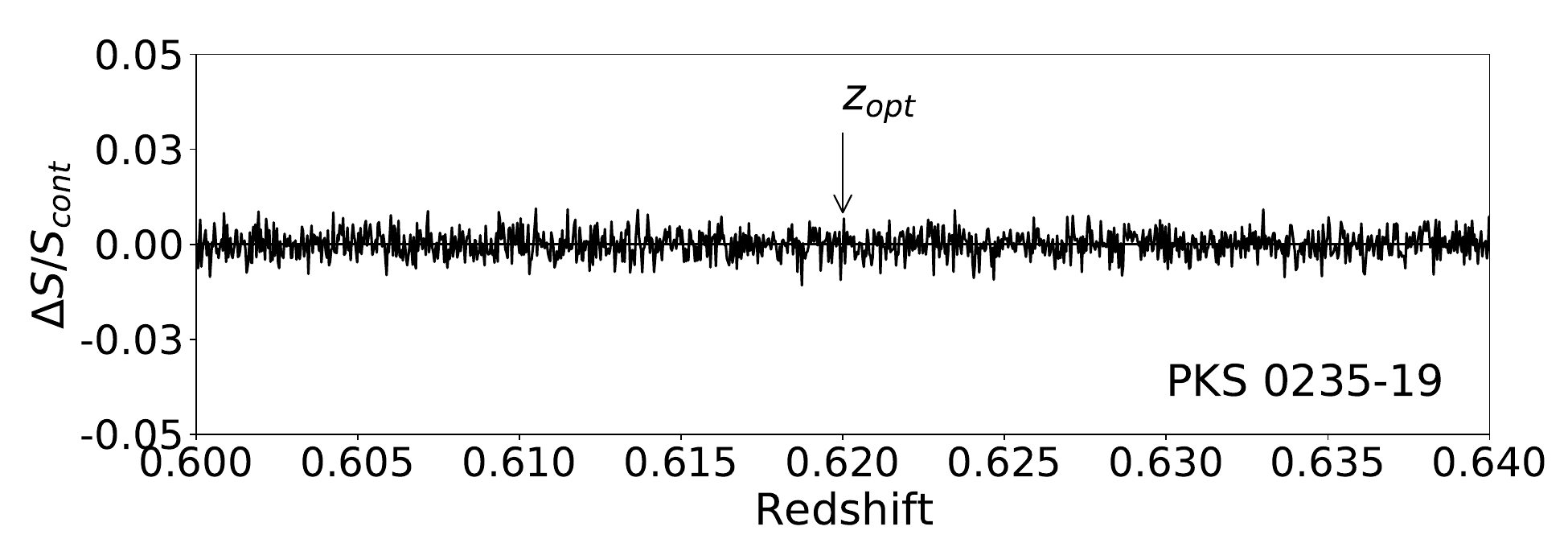}}
\end{minipage}
\vspace{-0.13cm}
\begin{minipage}{0.48\linewidth}
\centering{\includegraphics[width=\linewidth]{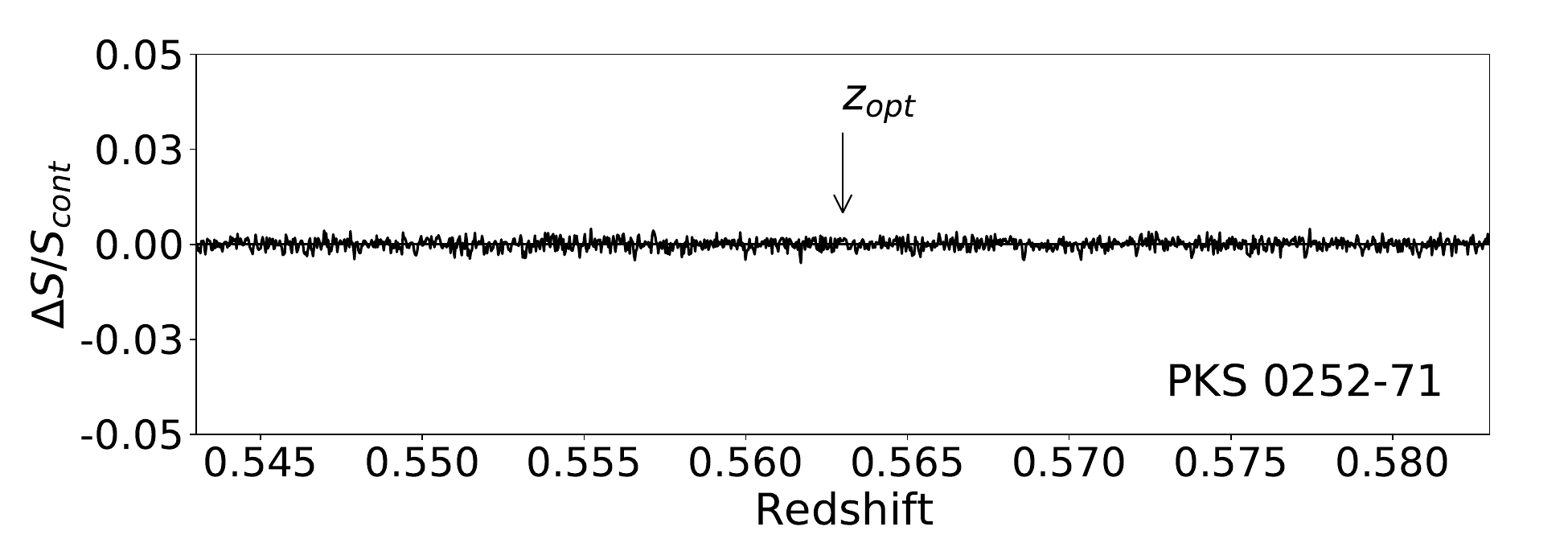}}
\end{minipage} 
\begin{minipage}{0.48\linewidth}
\centering{\includegraphics[width=\linewidth]{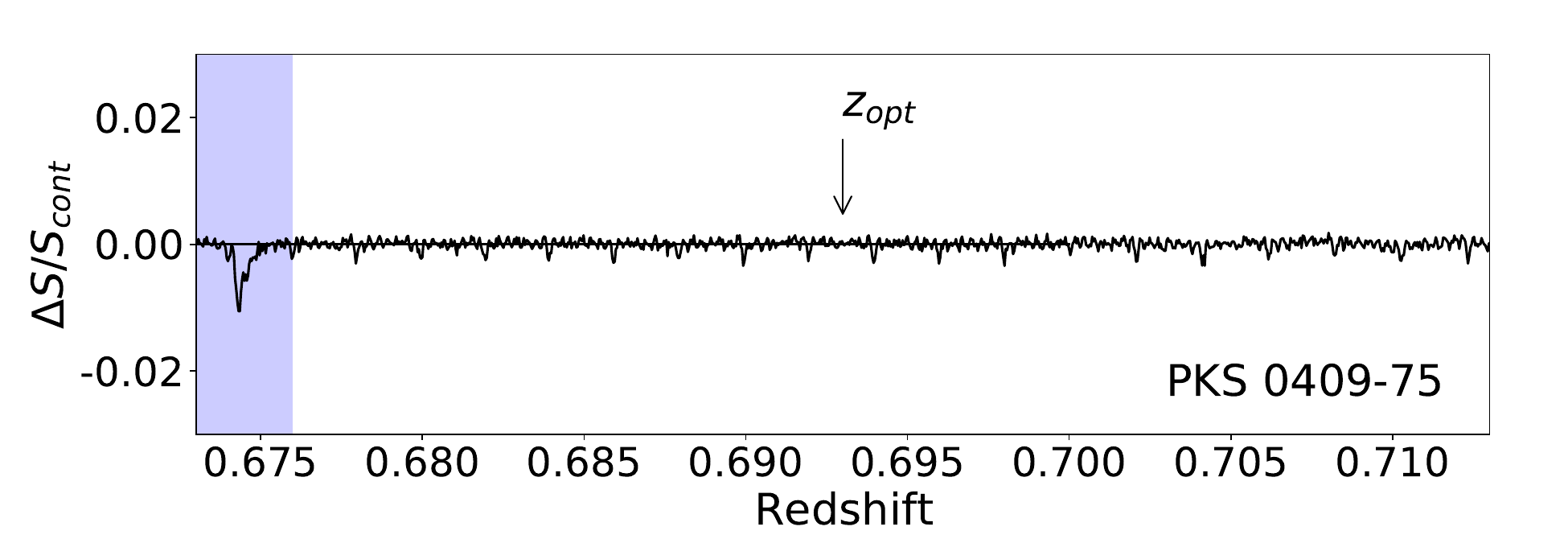}}
\end{minipage}
\vspace{-0.13cm}
\begin{minipage}{0.48\linewidth}
\centering{\includegraphics[width=\linewidth]{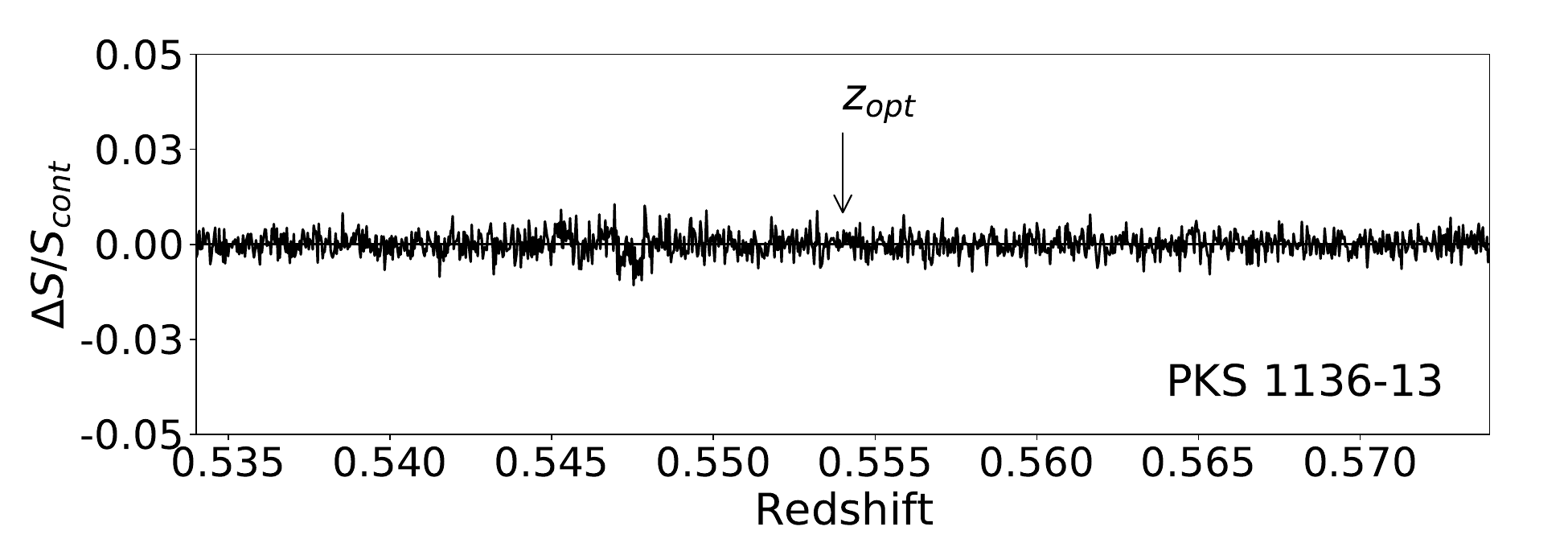}}
\end{minipage}
\begin{minipage}{0.48\linewidth}
\centering{\includegraphics[width=\linewidth]{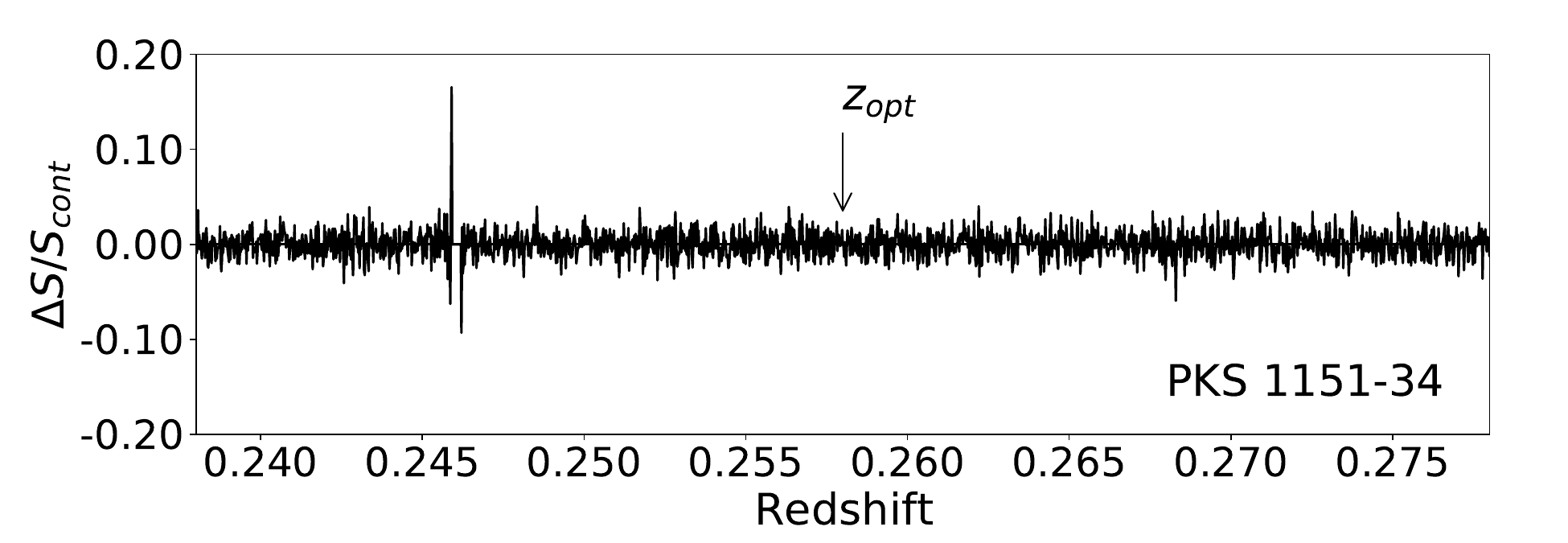}}
\end{minipage}
\vspace{-0.13cm}
\begin{minipage}{0.48\linewidth}
\centering{\includegraphics[width=\linewidth]{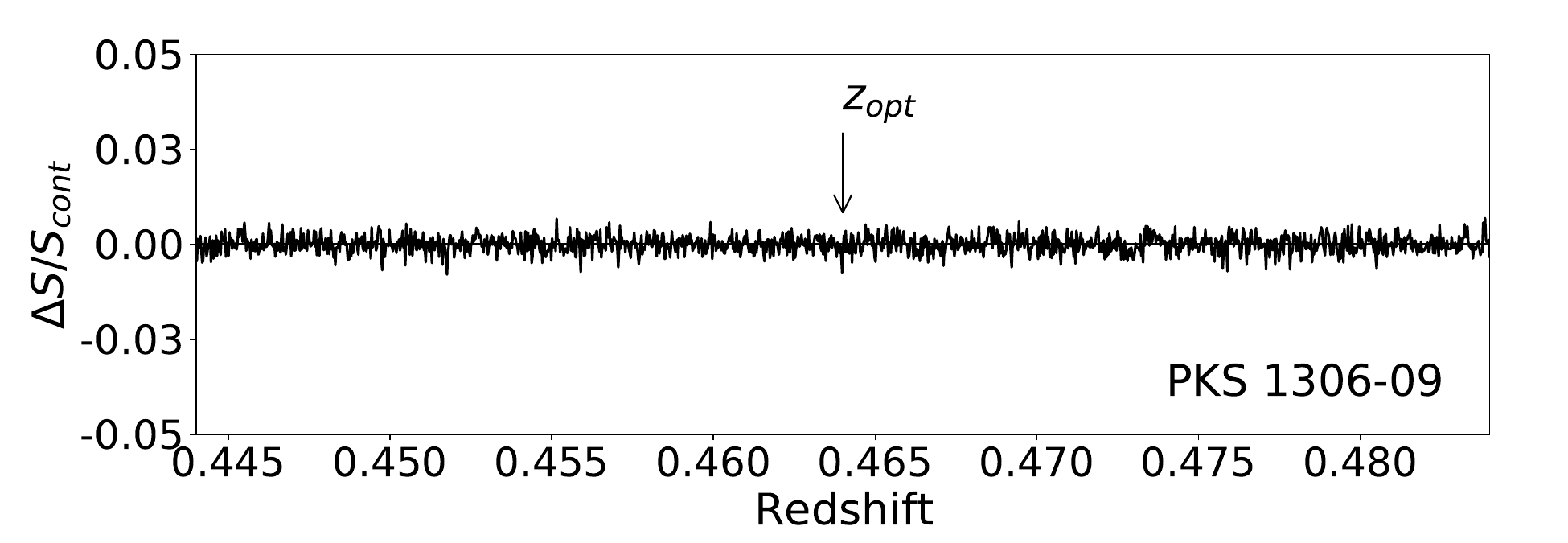}}
\end{minipage}
\begin{minipage}{0.48\linewidth}
\centering{\includegraphics[width=\linewidth]{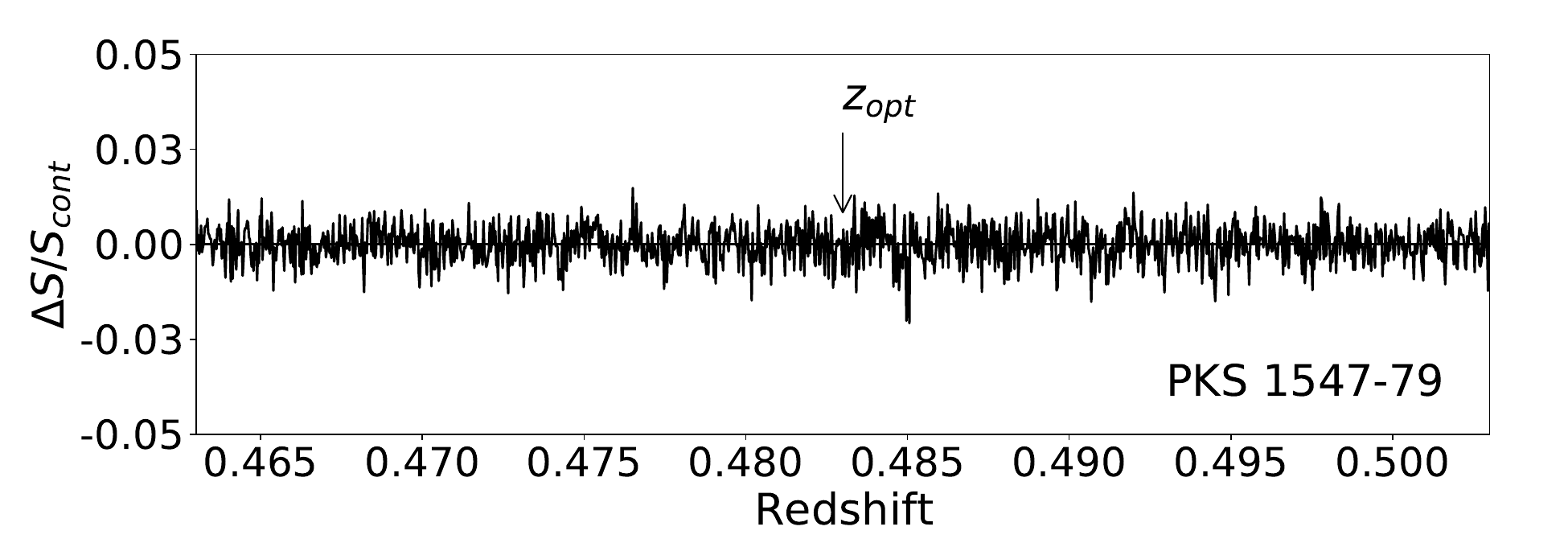}}
\end{minipage}
\vspace{-0.13cm}
\begin{minipage}{0.48\linewidth}
\centering{\includegraphics[width=\linewidth]{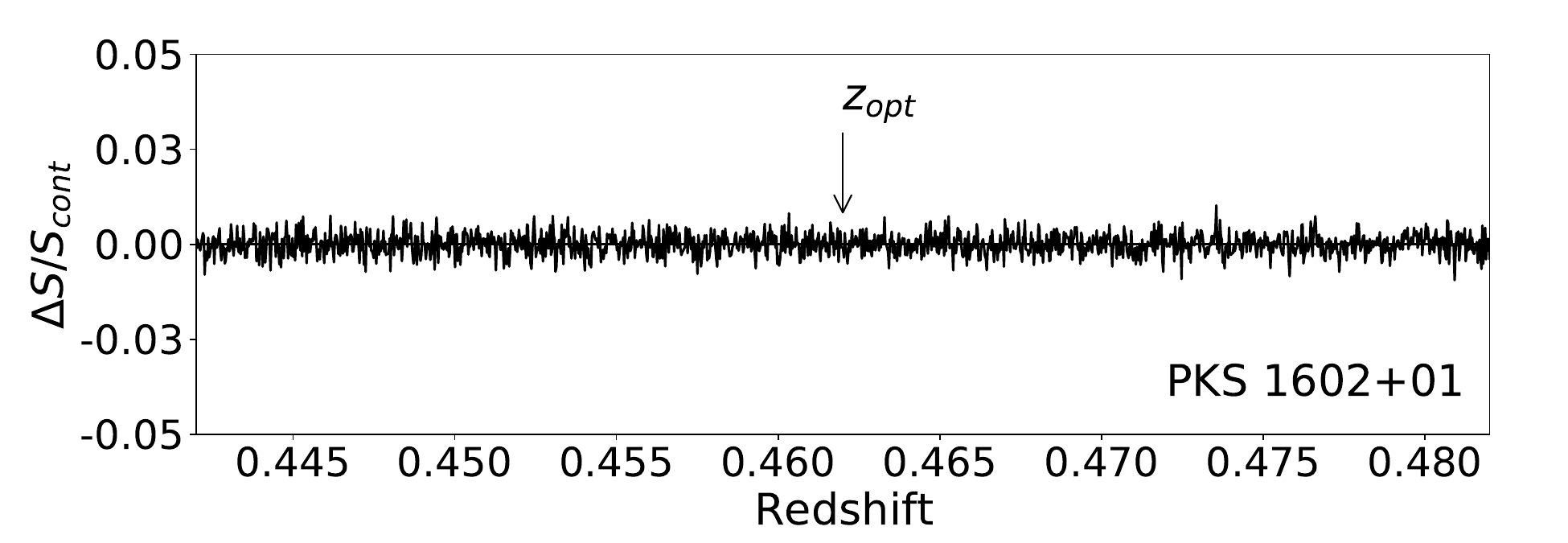}}
\end{minipage}
\begin{minipage}{0.48\linewidth}
\centering{\includegraphics[width=\linewidth]{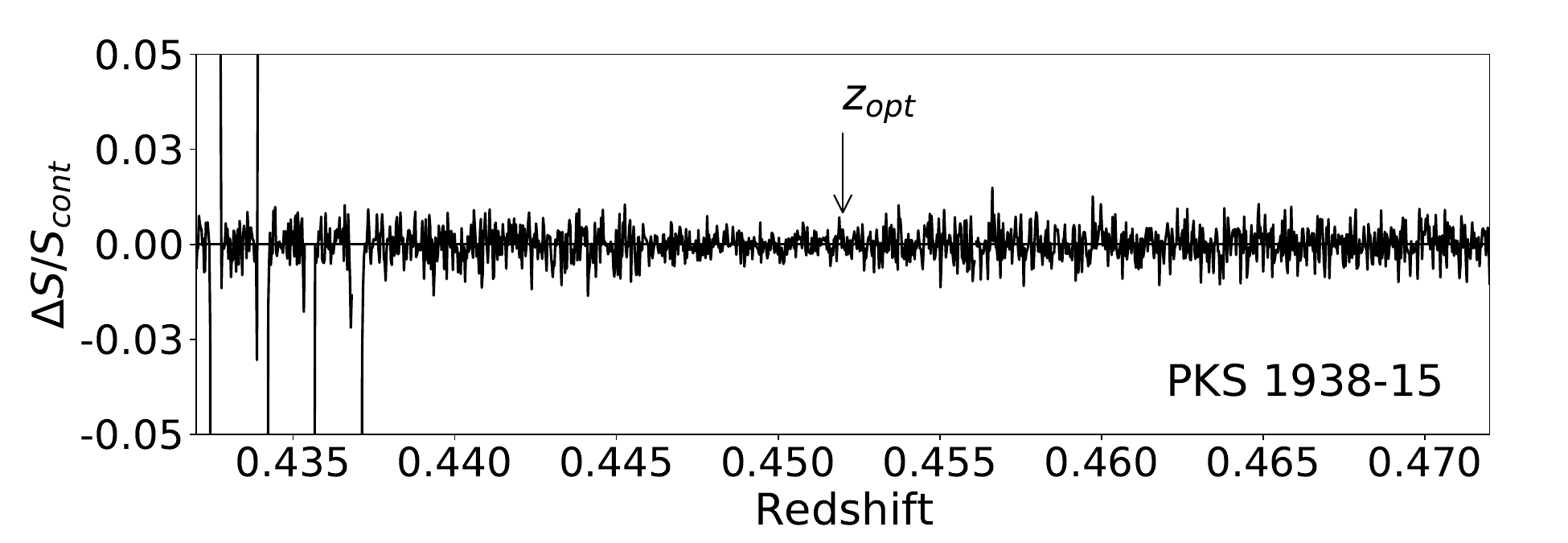}}
\end{minipage}
%\vspace{-0.4cm}
\begin{minipage}{0.48\linewidth}
\centering{\includegraphics[width=\linewidth]{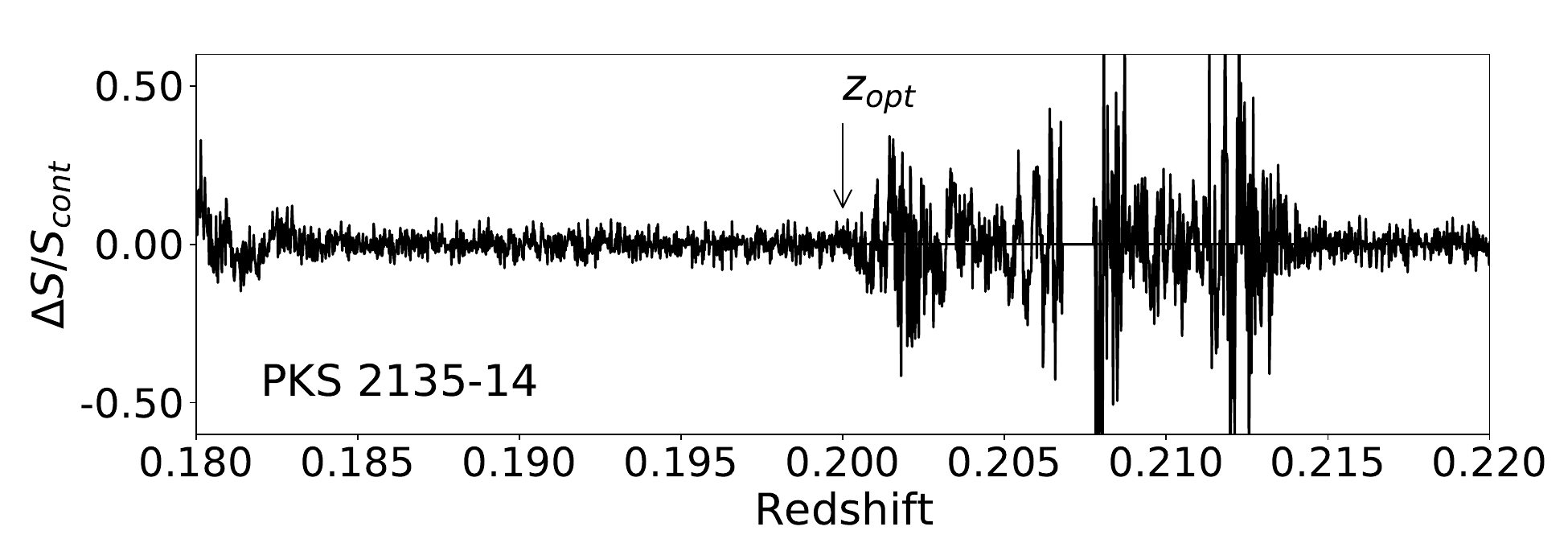}}
\end{minipage}
\begin{minipage}{0.48\linewidth}
\centering{\includegraphics[width=\linewidth]{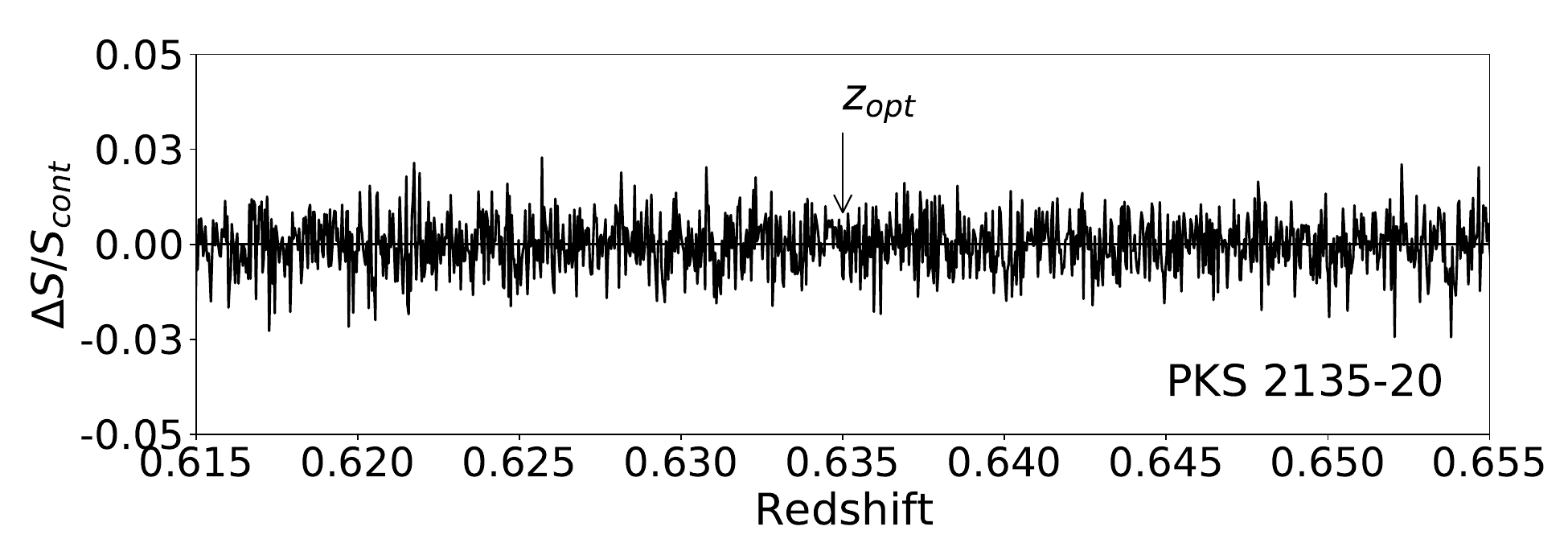}}
\end{minipage}
\caption{ASKAP-BETA spectra of 16 sources selected from the 2-Jy sample showing the fractional absorption (the measured flux density in absorption ($\Delta S$) divided by the continuum flux density ($S_{\rm cont}$)) for a small range around the optical redshift of the host galaxy (marked by the arrow). \mbox{H\,{\sc i}} 21-cm absorption is observed in PKS\,0023$-$26 at $z=0.322$ and PKS\,0409$-$75 at $z=0.674$, highlighted by the blue shaded boxes. Small artefacts can be seen in the spectrum of PKS\,0409-75 corresponding to a channelisation error at the edges of the 1\,MHz coarse channels. This issue was discovered during the ASKAP commissioning period and has since been corrected. The impact of RFI from aircraft and satellites is much greater at the higher frequencies (corresponding to $z_{\mbox{H\,{\sc i}}}<0.4$). \label{allspectra}}
\end{figure*}

\mbox{H\,{\sc i}} 21-cm absorption was detected in two of the sixteen targets observed with ASKAP; one associated with PKS\,0023$-$26 at a redshift of $z=0.322$ \citep{Vermeulen2003} and a new detection towards PKS\,0409$-$75 at a redshift of $z=0.674$. The \mbox{H\,{\sc i}} spectra are shown in Figure \ref{allspectra}. Here we display a small fraction of the total bandwidth around the expected frequency of the \mbox{H\,{\sc i}} 21-cm line determined from the optical redshift of the radio galaxy. An example of the full spectrum obtained from ASKAP-BETA is shown in \ref{fullspectrum}. The primary aim of this work is to search for associated absorption, but the wide bandwidth provided by ASKAP-BETA also allows us to search for any intervening absorption along the line-of-sight. No other detections of \mbox{H\,{\sc i}} 21-cm absorption were observed in this sample, but given the small sample size and amount of bandwidth flagged due to RFI at the lower frequencies this is not unexpected.

\subsection{\mbox{H\,{\sc i}} properties of the sample}

While this paper presents a comparatively small sample of 16 targets that were searched for \mbox{H\,{\sc i}} 21-cm absorption, a detection rate of 12.5\% (2/16) may appear to be at the lower end of the range of previously reported detection rates for associated 21-cm absorption \citep{Morganti2001, Vermeulen2003, Curran2008, Chandola2013, Gereb2015, Maccagni2017, Dutta2018, Murthy2021}. However, a number of uncertainties make it difficult to confirm this and carry out a fair comparison with previous studies. The biggest factors contributing to this are: (1) the wide range in integration time (and therefore sensitivity reached) for objects in the sample as all observations were carried out in an opportunistic fashion during the commissioning period, and (2) the large range of radio continuum source structure of our sample. The latter leads to highly uncertain covering factors ($C_{\rm f}$) which can change the optical depth and \mbox{H\,{\sc i}} column density limits by orders of magnitude. 

The covering factor is often estimated by using the ratio of core to total flux density (`$R$') which provides a measure of the compactness of the radio source, limited by the spatial resolution of the available data (see e.g. \citealt{Kanekar2009}). For compact sources (i.e. such as those classified as CSS in Table \ref{sampletab}) this gives a covering factor of $C_{\rm f}=1$ which is the typically assumed $C_{\rm f}$ for most 21\,cm absorption line studies, driven by the fact that most samples are selected based on source compactness, or a minimum core flux density (e.g. \citealt{Morganti2001, Maccagni2017}). In addition, we also need to make an assumption about $T_{\rm spin}$, the harmonic mean spin (excitation) temperature over the line-of-sight gas. At $z < 1$, the measured spin temperatures on sight lines through galaxies range from 100-1000\,K \citep{Kanekar2014, Allison2021} so in the absence of a direct measurement for these sources we assume a spin temperature of $T_{\rm spin}=500$\,K. This value was chosen as the approximate mid-point of the range of observed $T_{\rm spin}$, but could easily change the estimated $N_{\rm HI}$ by a factor of 2 if $T_{\rm spin}$ is similar to the Milky Way value of $\sim 300$\,K \citep{Dickey2009, Murray2018}, or $\sim 1000$\,K as observed in some damped Lyman-$\alpha$ systems (DLAs, e.g. \citealt{Kanekar2009b, Kanekar2014}) and in the central regions of AGN \citep{Bahcall1969, Holt2006}.

Having measured the absorption fraction ($\frac{\Delta S}{S_{\rm cont}}$), and using these assumptions for $C_{\rm f}$ and $T_{\rm spin}$ we can calculate the optical depth ($\tau$) and $N_{\rm HI}$ from the following equations:

\begin{equation}
\tau = -\ln \left(1 - \frac{\Delta S}{C_{\rm f} S_{\rm cont}}\right) \label{tau}.
\end{equation}

\begin{equation}
N_{\rm HI} = 1.823\times10^{18}  T_{\rm spin} \int \tau d\nu \label{nhi}.
\end{equation}

In Table \ref{tautab} we calculate $N_{\rm HI}$ using both $C_{\rm f}=1$ and $C_{\rm f}=R$ where this information was available. The core-total flux density ratios ($R$) were calculated from the higher resolution 4.8\,GHz radio data observed with either the VLA or ATCA as given in Table \ref{sampletab} \citep{Morganti1993}. Assuming a source covering factor of 1 is reasonable for compact radio sources, but is unphysical for extended radio sources with little core flux and will significantly underestimate $N_{\rm HI}$. On the other hand, our detection of \mbox{H\,{\sc i}} absorption towards PKS\,0409$-$75 demonstrates that using the core-total flux density ratio is also not always representative of the true covering factor (see Section \ref{pks0409HIproperties}). 

\begin{table}
\caption{Fractional absorption ($\Delta S / S_{\rm cont}$) and \mbox{H\,{\sc i}} column density measurements for 2-Jy sources observed with ASKAP-BETA. For sources where no \mbox{H\,{\sc i}} absorption was detected we report 3 sigma upper limits per channel and assume a velocity width of 30\,km\,s$^{-1}$ to calculate an upper limit for $N_{\rm HI}$. Due to the large uncertainties in the covering factor for this sample we present two values of $N_{\rm HI}$ to indicate the range of column densities probed by these observations depending on the assumed $C_{\rm f}$. A spin temperature of $T_{\rm spin}=500$\,K is used throughout.  \label{tautab}}
%\begin{center}
\begin{threeparttable}
\begin{tabular}{lcccc}
\hline
%%VERSION 3: Assume Tspin=500K, Cf=1 
Source & $R$ & $\frac{\Delta S}{S_{\rm cont}}$ & $N_{\rm \mbox{H\,{\sc i}}}$ ($C_{\rm f}$=$R$) & $N_{\rm \mbox{H\,{\sc i}}}$ ($C_{\rm f}$=1) \\
&&& $(cm^{-2})$ & $(cm^{-2})$ \\
\hline
PKS\,0023$-$26 & 1.0 & 0.0122 & $1.49\times10^{21}$ & $1.49\times10^{21}$ \\%& \\
PKS\,0035$-$02 & 0.335 & $<$0.050 & $<4.1\times10^{21}$ & $<1.37\times10^{21}$ \\
PKS\,0039$-$44 & - & $<$0.050 & - & $<1.37\times10^{21}$ \\%& $<2.76\times10^{18}$\\
PKS\,0105$-$16 & 0.002 & $<$0.014 & $<1.91\times10^{23}$ & $<3.83\times10^{20}$ \\%& $<7.71\times10^{17}$\\
PKS\,0117$-$15 & - & $<$0.009 & - & $<2.46\times10^{20}$ \\%& $<5.09\times10^{17}$\\
PKS\,0235$-$19 & 0.0001 & $<$0.009 & $<2.46\times10^{24}$ & $<2.46\times10^{20}$ \\%& $<5.09\times10^{17}$\\
PKS\,0252$-$71 & 1.0 & $<$0.010 & $<2.73\times10^{20}$ & $<2.73\times10^{20}$ \\%& $<5.41\times10^{17}$\\
PKS\,0409$-$75$^*$ & - & 0.0107 & - & $2.16\times10^{21}$\\%& \\
PKS\,1136$-$13 & - & $<$0.0075 & - & $<2.05\times10^{20}$ \\%& $<4.10\times10^{17}$\\
PKS\,1151$-$34 & 1.0 & $<$0.029 & $<7.93\times10^{20}$ & $<7.93\times10^{20}$ \\%& $<1.59\times10^{18}$\\
PKS\,1306$-$09 & 1.0 & $<$0.0063 & $<1.72\times10^{20}$ & $<1.72\times10^{20}$ \\%& $<3.45\times10^{17}$\\
PKS\,1547$-$79 & 0.002 & $<$0.013 & $<1.78\times10^{23}$ & $<3.55\times10^{20}$ \\%& $<7.22\times10^{17}$\\
PKS\,1602$+$01 & 0.061 & $<$0.0078 & $<3.50\times10^{21}$ & $<2.13\times10^{20}$ \\%& $<4.27\times10^{17}$\\
PKS\,1938$-$15 & - & $<$0.012 & - & $<3.28\times10^{20}$ \\%& $<6.73\times10^{17}$\\
PKS\,2135$-$14 & 0.096 & $<$0.093 & $<2.65\times10^{22}$ & $<2.54\times10^{21}$ \\%& $<5.09\times10^{18}$\\
PKS\,2135$-$20 & 1.0 & $<$0.021 & $<5.74\times10^{20}$ & $<5.74\times10^{20}$ \\%& $<1.13\times10^{18}$\\
\hline
\end{tabular}
\begin{tablenotes}
\item$^*$ A source covering factor of $C_{\rm f}=0.3$ was assumed for this source as discussed in Section \ref{pks0409HIproperties}.
\end{tablenotes}
\end{threeparttable}
%\end{center}
\end{table}

Although the average spectral noise reached in these observations is similar to that achieved in similar studies using ASKAP-BETA (e.g. \citealt{Allison2015, Moss2017, Glowacki2019, Sadler2020}), these studies primarily focused on compact radio sources and therefore assumed a uniform source covering factor, similar to the majority of \mbox{H\,{\sc i}} absorption searches carried out with other radio telescopes. Due to the large uncertainties associated with estimating the covering factor when including resolved or highly extended radio sources from the 2-Jy sample, it is challenging to carry out a fair comparison on the detection rate in this sample with previous studies. In the remainder of this paper we focus on the two detections of \mbox{H\,{\sc i}} absorption towards PKS\,0023$-$26 and PKS\,0409$-$75. 

\subsection{\mbox{H\,{\sc i}} absorption in PKS\,0023$-$26}

PKS\,0023$-$26 is a compact steep spectrum radio source at $z=0.32188 \pm 0.00004$ \citep{Santoro2020} with a linear size of 1.97\,kpc \citep{Tzioumis2002}. The radio source is identified with an $m_v=19.5$ galaxy \citep{WallPeacock1985} and deep optical and IR follow-up have found evidence for active star formation \citep{Tadhunter2002, Holt2007, Dicken2012}. Recent ALMA observations of this source also reveal that it is extremely rich in molecular gas \citep{Morganti2021}.

\begin{figure}
\centering{\includegraphics[width=\linewidth]{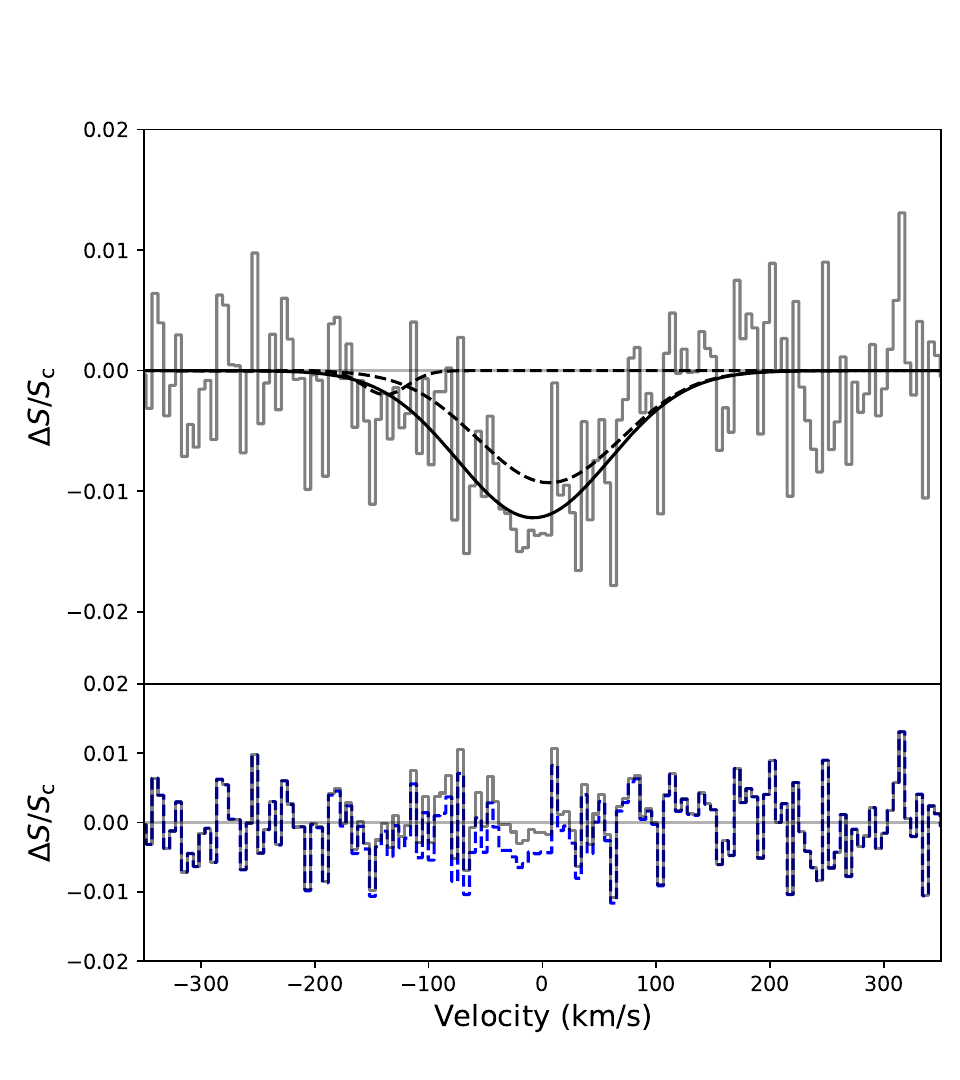}}
\caption{{\it Top:} \mbox{H\,{\sc i}} absorption detected in PKS\,0023$-$26 where the velocity-axis is defined with reference to the optical redshift $z=0.32188$. The solid line shows the best fitting parameters of a single Gaussian component with FWHM = 133.5 km\,s$^{-1}$ and peak optical depth $\tau = 0.0122$. The dashed lines show the line parameters, and the additional component reported by \citet{Vermeulen2003}, shifted to the same systemic velocity for comparison. {\it Bottom:} The residual spectrum after subtraction of the best fitting parameters determined by FLASHfinder (solid grey line) and the residual spectrum after subtraction of the parameters reported in \citet{Vermeulen2003} (blue dashed line). \label{pks0023_linefinder}}
\end{figure}

Observations with ASKAP-BETA reveal \mbox{H\,{\sc i}} absorption at 1074.4\,MHz. To derive the properties of the \mbox{H\,{\sc i}} line we used FLASHFinder\footnote{https://github.com/drjamesallison/flash\_finder}, an automated line finding and characterisation tool, full details of which are described in \citet{Allison2012}. A single Gaussian component with a rest-frame FWHM$=133.5\substack{+18.9 \\ -16.5}$ km\,s$^{-1}$, peak optical depth $\tau = 0.0122\pm 0.0014$ and redshift $z_{\rm HI}=0.32184\pm0.00003$ provided the best fit to the data. This best fitting model is shown in Fig \ref{pks0023_linefinder} where the systemic velocity has been set to the optical redshift of $z=0.32188$ \citep{Santoro2020}. Since PKS\,0023$-$26 is a known CSS source (with Very Long Baseline Interferometry (VLBI) imaging confirming a linear extent $<2$\,kpc; \citealt{Tzioumis2002}) we assume a covering factor of 1 and a spin temperature of 500\,K to obtain a \mbox{H\,{\sc i}} column density of 
\begin{equation}
N_{\rm HI} = 1.49\times10^{21} {\left(\frac{T_{\rm{spin}}}{500\rm{K}}\right)}{\left(\frac{C_{\rm f}}{1.0}\right)}^{-1} \rm{cm}^{-2}.
\end{equation}

\mbox{H\,{\sc i}} 21-cm absorption was previously detected in this source by \citet{Vermeulen2003} using the Westerbork Synthesis Radio Telescope (WSRT). Due to the larger sensitivity of the WSRT array compared to ASKAP-BETA, these observations revealed two components of the absorption line: a deeper component centred close to the systemic velocity of the galaxy ($v=-30$\,km\,s$^{-1}$, $\Delta v=126$\,km\,s$^{-1}$ with peak optical depth $\tau=0.0093$) and a second, shallower component blueshifted from the systemic ($v=-174$\,km\,s$^{-1}$, $\Delta v=39$\,km\,s$^{-1}$ with peak optical depth $\tau=0.002$)\footnote{We note that the rest-frame velocities measured in the WSRT observations are defined from the most accurate redshift available at the time of publication of $z=0.322$ \citep{DiSeregoAlighieri1994}}. 

Figure \ref{pks0023_linefinder} overlays these line fits to visually compare with the single Gaussian component identified in the ASKAP-BETA spectrum. The shallower component is not recovered in the ASKAP-BETA observations due to the higher noise levels, but while the velocity width is similar, the deeper component shows a slight change in peak optical depth between the WSRT and ASKAP-BETA observations. This is mostly clearly seen in the residual spectra after subtracting these two different fits as shown in the bottom panel of Figure \ref{pks0023_linefinder}. The uncertainties associated with the peak optical depth measurements indicate they are within 2 sigma and therefore any possible variability is unlikely to be statistically significant. To further verify this, the line parameters reported by \citet{Vermeulen2003} were subtracted from the ASKAP-BETA spectrum and the residual spectrum run through the FLASHfinder line-finding tool to search for evidence of an absorption feature remaining. No significant lines were detected confirming that the measured line parameters are consistent with each other within the noise. 

\subsection{\mbox{H\,{\sc i}} absorption in PKS\,0409$-$75}

PKS\,0409$-$75 is a powerful FRII radio galaxy with a projected linear size of 85\,kpc \citep{Large1981, Morganti1999}\footnote{PKS\,0409$-$75 is also sometimes referred to as PKS\,0410$-$75 (e.g. in the NASA Extragalactic Database; https://ned.ipac.caltech.edu/ )}. The host galaxy is a magnitude m$_V$ = 21.6 galaxy which is resolved into two components and displays extended emission in [OII]$\lambda$3727 \citep{DiSeregoAlighieri1994}. Fitting stellar population models to the optical spectrum, \citet{Holt2007} found that PKS\,0409$-$75 has a very young stellar population with an age of approximately 0.02 Gyr. The extended emission in [OII], a young stellar population, and complicated optical structure, point towards recent merger activity in this galaxy.

\begin{figure}
\centering{\includegraphics[width=\linewidth]{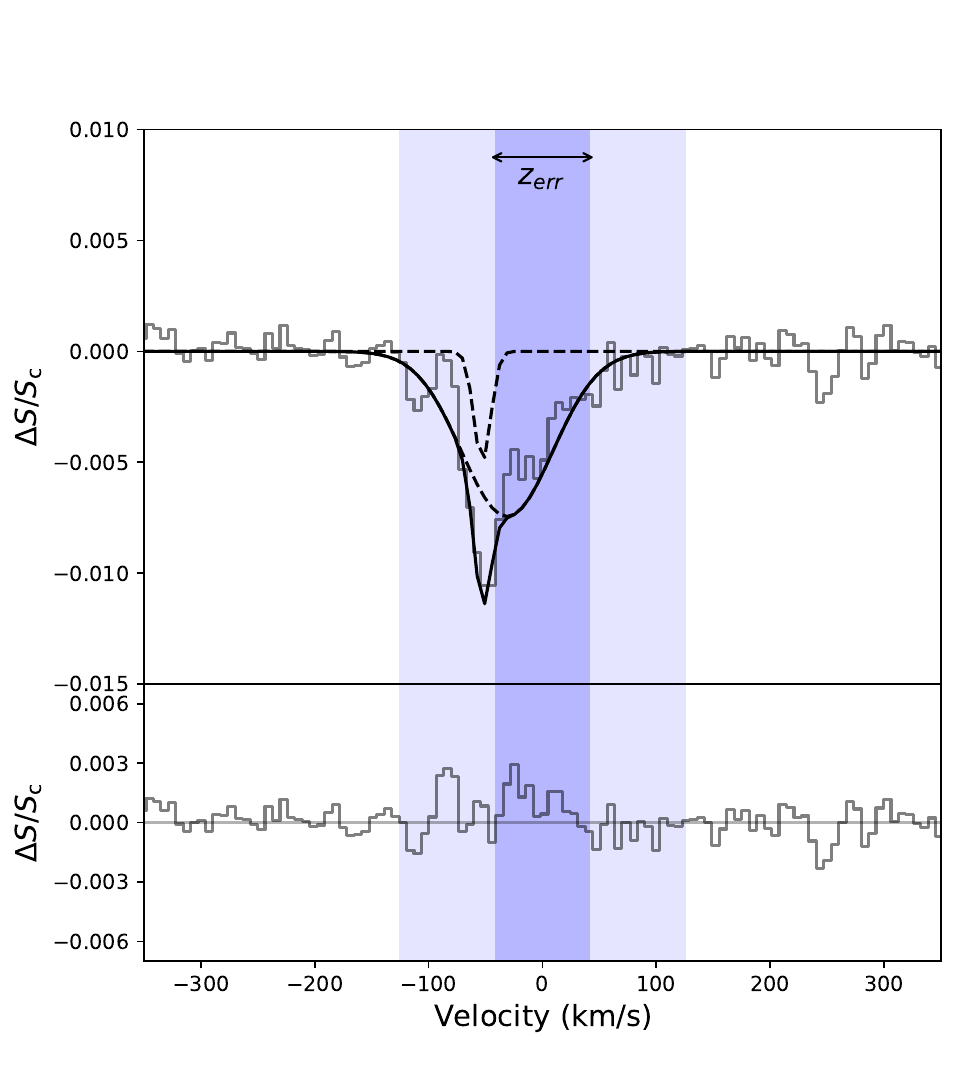}}
\caption{{\it Top:} Spectrum and best fitting model to the \mbox{H\,{\sc i}} absorption line detected towards PKS\,0409$-$75. The dashed lines plot the individual Gaussian components and the solid line shows the combined fit. {\it Bottom:} Residual spectrum after subtraction of the best fitting parameters shown in the top panel. The velocity axis is given with respect to the optical redshift shown in Section \ref{optical0409} with the shaded regions showing the 1- and 3-$\sigma$ errors ($z_{\rm opt} = 0.6746 \pm 0.00024$). \label{pks0409_linefinder}}
\end{figure}

ASKAP-BETA observations revealed \mbox{H\,{\sc i}} absorption detected along the line-of-sight to PKS\,0409$-$75, at a frequency of 848.5\,MHz. Using the FLASHfinder line-fitting tool the absorption line is best fit by two Gaussian components as shown in Figure \ref{pks0409_linefinder}; a narrow line with FWHM $=14.8\substack{+1.1 \\ -0.9}$\,km\,s$^{-1}$, peak redshift $z=0.67431\pm0.00005$ and peak optical depth $0.0050\pm0.0003$ and a broader component with FWHM $=80.1\substack{+2.2 \\ -1.9}$\,km\,s$^{-1}$, redshift $z=0.67443\pm0.00005$ and peak optical depth $0.0074\pm0.0001$. However, the most striking feature of this detection is that the \mbox{H\,{\sc i}} absorption is at a redshift of $z_{\rm \mbox{H\,{\sc i}}}$ = 0.674, offset from the systemic velocity of the host galaxy (determined by the optical redshift of $z=0.693$) by more than $-3300$\,km\,s$^{-1}$.

Such an extreme velocity offset implies that the neutral gas detected via the 21-cm absorption signal is not associated with PKS\,0409$-$75 itself, but more likely originates from a nearby galaxy in the surrounding environment of this powerful radio source. Although there is some evidence for merger-related activity in the host galaxy of PKS\,0409$-$75 \citep{RamosAlmeida2011}, the velocity offset is significantly larger than what is typically associated with tidal features. Tidal features associated with individual galaxies or galaxy pairs are typically less than 200\,km\,s$^{-1}$ \citep{Bekki2005,Koribalski2020} while the spread of tidal debris in galaxy groups or clusters can be larger, up to $\sim500$\,km\,s$^{-1}$ and $\sim1000$\,km\,s$^{-1}$ respectively \citep{Saponara2018,Lee-Waddell2019,Namumba2021}. A velocity difference of more than 3000\,km\,s$^{-1}$ has not been observed and is unlikely in a bound system. 

Previous deep optical imaging of PKS\,0409-75 presented in \citet{RamosAlmeida2011} shows evidence for a separate galaxy co-incident with the southern radio lobe, however, no spectroscopic information was available to indicate if this galaxy is at the same redshift as the \mbox{H\,{\sc i}} absorption line. 

\section{Optical follow-up of PKS\,0409$-$75}

To verify that this galaxy observed in previous optical images could be a possible association of the H{\sc i} feature, additional imaging and longslit spectroscopy were obtained using the Gemini-South telescope. 

\subsection{Optical Imaging}

Follow-up multi-band imaging of PKS\,0409$-$75 was carried out on 2017 August 28 using the Gemini Multi-Object Spectrograph (GMOS) on the Gemini-South 8m telescope (ProgID number GS-2017B-Q-63). The 5.5 arcmin field was centred on PKS\,0409$-$75 and a 300 second exposure time used in each of the $g',r'$ and $i'$ filters with the aim of reaching limiting magnitudes of approximately $g'$=23.9, $r'$=23.5 and $i'$=23.3. The total exposure time in each band was split into 4$\times$75 second individual exposures to aid in cosmic ray rejection. The seeing conditions during the observations ranged from 0.7 -- 0.8\,arcsec. 

The data were reduced using the standard Gemini \textsc{iraf} packages and a gri-colour image made using the \textsc{astropy} and \textsc{aplpy} packages \citep{astropy2018, aplpy2019}. No standard stars were observed during the imaging observations, so the median photometric zero-points measured for each GMOS-S filter\footnote{https://www.gemini.edu/instrumentation/gmos/calibrations\#PhotStand} were used when calculating optical magnitudes. 

\subsection{Longslit Spectroscopy}

Longslit observations of PKS\,0409$-$75 were taken with the Gemini Multi-Object Spectrograph (GMOS) on the Gemini-South telescope on 2017 September 16 as part of the same observing programme (GS-2017B-Q-63). In order to distinguish between the two possible optical associations detected in previous Gemini imaging we orientated a 1.5\,arcsec slit at a position angle of 133\,deg to obtain spectra of both the host galaxy of PKS\,0409$-$75 and the nearby galaxy at the position of the eastern radio lobe.

The observations were taken using the R400 grating with central wavelengths of 700 and 705\,nm which allowed for dithering across the CCD chip gaps to obtain continuous spectral coverage from 490-910\,nm. Using 2x2 binning resulted in pixel sizes of 0.16\,arcsec spatially and 0.15\,nm in the spectral direction. 6x900s exposures were taken for each central wavelength leading to a total on source integration time of 3\,hr. GCALflat observations were taken in between the science observations while bias images and CuAr arc exposures were taken at the end of the night. 

Standard Gemini \textsc{iraf} tasks were used for bias subtraction, flat-fielding, wavelength calibration and sky subtraction for each exposure independently. The 2D spectra for each configuration were then combined to increase the S/N before extracting two apertures for each combined 2D spectrum; one at the position of the host galaxy of PKS\,0409$-$75 (denoted `galaxy A' hereafter) and another at the position of the galaxy at the position of the eastern radio lobe (`galaxy B'). The data were flux calibrated using an observation of a standard star (CD-34 241) taken on 2017 August 21. No telluric calibration was performed meaning the atmospheric A-band is still evident in the resulting spectra. Once calibrated, the 1D spectra were combined using the {\sc iraf scombine} task.

The reduced spectra of both galaxies were then run through the MARZ web-based redshifting software \citep{Hinton2016} to identify the redshift of both sources.

\subsection{Optical properties of PKS\,0409$-$75} \label{optical0409}

\begin{figure}
\centering{\includegraphics[width=\linewidth]{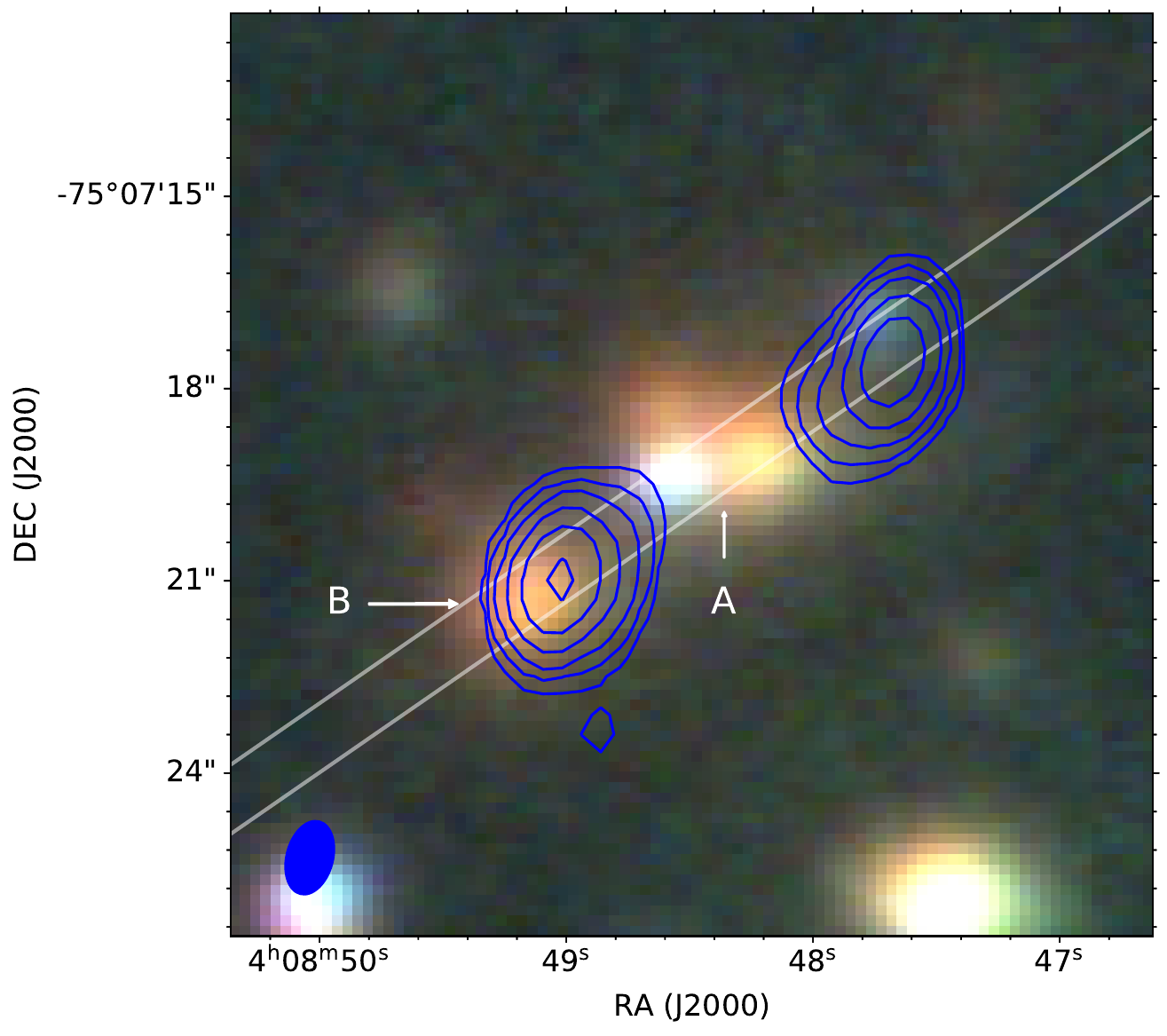}}
\caption{Gemini GMOS $gri$-image of PKS\,0409$-$75 and surrounding galaxies. The double nucleus in PKS\,0409$-$75 reported in \citet{RamosAlmeida2011} is clearly visible (galaxy A), as is a possible companion galaxy to the south-east (galaxy B). Radio contours from higher resolution 8 GHz ATCA observations are overlaid in blue, with the synthesised beam shown in the bottom left corner. The grey lines mark the position of the slit used in the Gemini-GMOS longslit observations discussed in Section \ref{optical0409}. Standard image orientation is used with North up and East to the left. \label{gemini_image}}
\end{figure}

A 3-colour image compiled from $g'$,$r'$ and $i'$ band observations is shown in Figure \ref{gemini_image} overlaid with contours from 8\,GHz Australia Telescope Compact Array (ATCA) observations \citep{Morganti1999}, which shows an elliptical galaxy at the position of the southern radio lobe. For ease of reference we denote the host galaxy of PKS\,0409$-$75 `galaxy A' and the galaxy coincident with the southern radio lobe `galaxy B' hereafter. We also note that the radio contours shown in Figure \ref{gemini_image} have much higher spatial resolution due to the higher observed frequency and longer baselines provided by the ATCA, resulting in a synthesised beam of $0\farcs7\times1\farcs2$ compared to the $\sim1\farcm5$ resolution provided by ASKAP-BETA. The continuum source is unresolved in the ASKAP-BETA observations presented here and therefore we cannot directly determine if the absorption line is detected along the line of sight to one of the radio lobes from the ASKAP-BETA data alone. 

\begin{figure*}
\centering{\includegraphics[width=\linewidth]{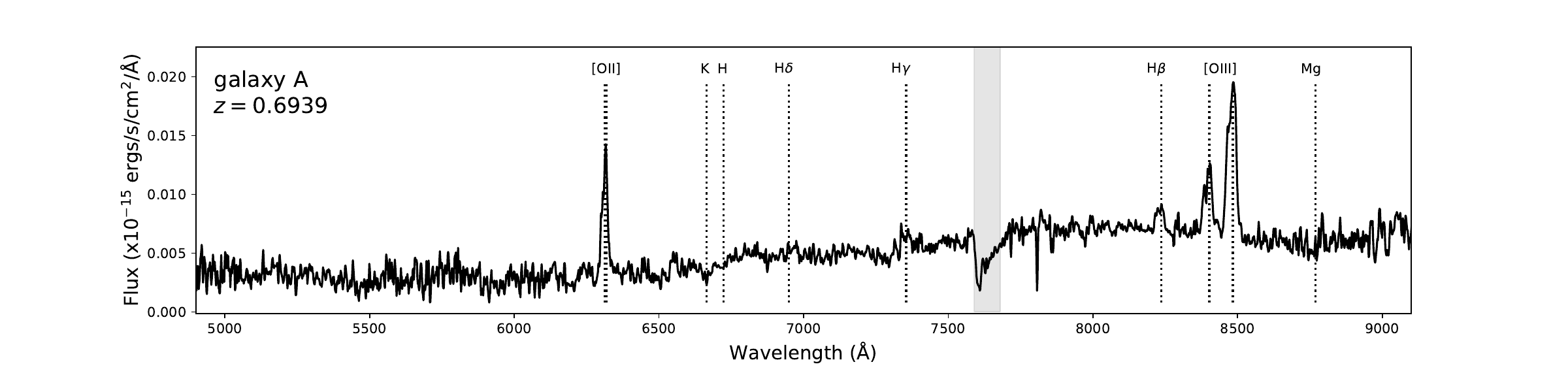}}
\centering{\includegraphics[width=\linewidth]{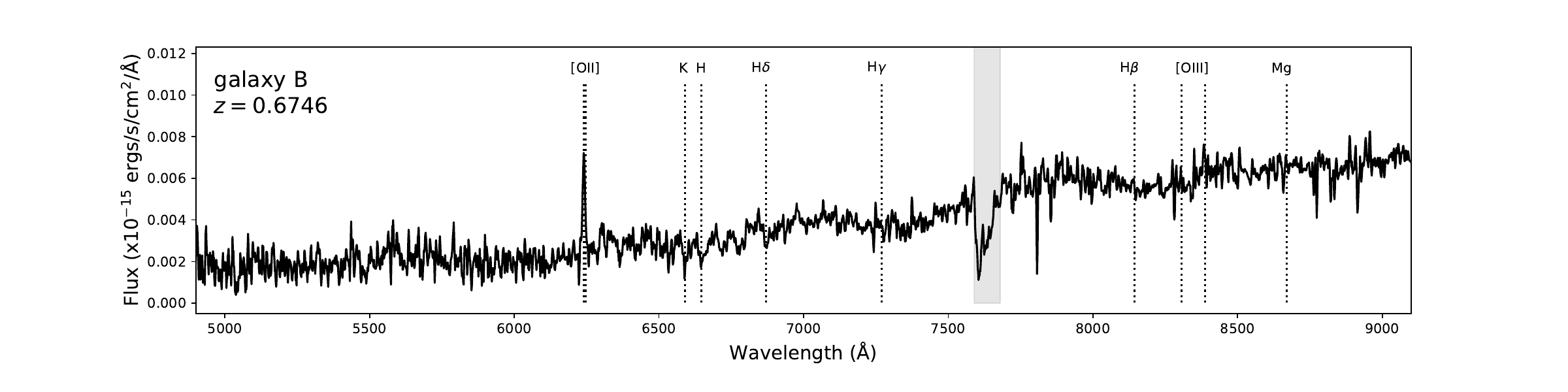}}
\caption{Gemini-South GMOS longslit spectra of PKS\,0409$-$75. The top panel shows the spectrum for the host galaxy (galaxy A in Fig \ref{gemini_image}) and the bottom panel shows the spectrum of galaxy B. Both spectra are shown in the observed frame with the dashed lines marking commonly observed lines at the redshift of each source ($z=0.6939$ and $z=0.6746$ for sources A and B respectively) and show where these features would lie at the observed redshift. Not all of these features are necessarily evident in the spectrum. The grey shaded area corresponds to the A-band atmospheric absorption feature.  \label{pks0409_gemini}}
\end{figure*}

Figure \ref{pks0409_gemini} shows the spectra obtained from the Gemini-GMOS observations of galaxies A (top) and B (bottom). Both the optical imaging and spectroscopy confirm the complex nature of the host galaxy of PKS\,0409$-$75 (galaxy A), with the 3-colour (gri) image showing clear evidence for a double nucleus as previously shown in \citet{RamosAlmeida2011}. The spectrum of galaxy A is similar to that observed by \citet{DiSeregoAlighieri1994} and also shows extended [OII] and [OIII] emission as discussed in \citet{Holt2007}. 

The redder optical colours for galaxy B on the other hand suggest that this is more likely to be a early-type galaxy. The spectrum of galaxy B shows [OII] emission, as well as evidence for weak [OIII] emission and the CaII H and K absorption lines, placing the source at a redshift $z=0.6746 \pm 0.00024$. This is very similar to the redshift of the absorption line detected with ASKAP-BETA (within $\sim 50$\,km\,s$^{-1}$), confirming galaxy B as the association of the \mbox{H\,{\sc i}} 21-cm absorption line. 

\section{Discussion}

\subsection{Properties of the \mbox{H\,{\sc i}} absorbing galaxy towards PKS\,0409$-$75} \label{pks0409HIproperties}

Our targeted follow-up observations of this source at optical-wavelengths indicate that the \mbox{H\,{\sc i}} 21-cm absorption system detected with ASKAP-BETA is associated with galaxy B in front of the eastern radio lobe of PKS\,0409$-$75. Using the redshift of galaxy B ($z=0.6746 \pm 0.00024$) as the systemic velocity, Figure \ref{pks0409_linefinder} shows that the \mbox{H\,{\sc i}} feature detected is slightly blueshifted compared to the systemic velocity, but consistent within the redshift error of 42\,km\,s$^{-1}$ derived from the optical follow-up\footnote{The redshift error is estimated using the Penalized Pixel-Fitting (pPXF) software \citep{Cappellari2017}}. The fact that most of the 21-cm absorption is observed blueward of the systemic velocity could indicate that only a portion of the \mbox{H\,{\sc i}} in galaxy B is being detected in absorption. This is somewhat verified by the optical image shown in Figure \ref{gemini_image} which shows that galaxy B is not entirely in front of the background radio lobe of PKS\,0409$-$75, but higher resolution images in both the radio and optical are needed to confirm this. The peak of the radio emission is offset from the centroid of galaxy B meaning that an offset of $\sim 75$\,km\,s$^{-1}$ might be expected for a typical galaxy rotation velocity of $\sim 150$\,km\,s$^{-1}$ if this sight-line was probing only a small fraction of the galaxy \citep{Reeves2016}. The width of the \mbox{H\,{\sc i}} line is broader than typically observed for intervening galaxies where the mean of the distribution is $\sim$30\,km\,s$^{-1}$ \citep{Curran2016}, however this could be explained by the fact that diffuse continuum emission from the background radio lobe could be illuminating multiple sight-lines through galaxy B. We also looked for evidence of rotation along the slit, but as a larger aperture was needed to get sufficient S/N to detect the weak emission lines there was no clear indication of any rotation. 

To calculate the \mbox{H\,{\sc i}} column density we assume a covering factor of 0.3, roughly estimated  based on the flux ratio between the two radio lobes (the eastern lobe contains $\sim 60$\% of the total flux), combined with the fact that galaxy B intercepts approximately half of the radio lobe as shown in Figure \ref{gemini_image}. Again assuming a spin temperature $T_{\rm spin}=500$\,K and using the measured integrated optical depth of $0.71\pm0.01$,  Equations \ref{tau} and \ref{nhi} give a 21-cm \mbox{H\,{\sc i}} column density of 
\begin{equation}
N_{\rm HI}= 2.16\times10^{21} {\left(\frac{T_{\rm{spin}}}{500\rm{K}}\right)}{\left(\frac{C_{\rm f}}{0.3}\right)}^{-1} \rm{cm}^{-2}.
\end{equation}

We estimate a star-formation rate of 1.24\,$M_\odot$ yr$^{-1}$ using the Penalized Pixel-Fitting (pPXF) software \citep{Cappellari2017} to fit and measure the flux of the [OII] emission and following the Kennicutt relation \citep{Kennicutt1998}. Since the balmer lines were not detected in the longslit observations it was not possible to do any extinction correction when calculating the SFR so this value represents a rough estimate. It should also be noted that [OII] emission is not typically the best tracer to estimate the star-formation rate due to contamination of AGN or shocks which can also produce [OII] emission \citep{Maddox2018}. However, there is no strong evidence for AGN activity in galaxy B and the moderate amount of star-formation is consistent both with the \mbox{H\,{\sc i}} column density and typical SFR of early-type galaxies \citep{Serra2012,Davis2014}.

Using the {\sc SExtractor} source finding software \citep{sextractor}, we measure optical magnitudes of $g'=21.32 \pm0.11$, $r'=20.13\pm0.04$ and $i'=18.90\pm0.03$ using the {\sc automag} option for the absorbing galaxy (galaxy B). The $r'$-band image was used as the detection image for all filters to ensure that a common aperture was used for the source finding. These optical magnitudes were corrected for Galactic extinction using $A_{\lambda}$ values given by \citet{Schlafly2011} and K-corrected to give a restframe ($g'-r'$) colour of 1.12~mag using the method in \citet{Chilingarian2010}. Galaxy B was also detected in the VISTA Hemisphere Survey (VHS; \citealt{McMahon2013, vista}) with source name VHS J040849.12$-$750721.5 and magnitudes $J=17.10$ and $Ks=15.48$ mag. Using ($g'-r'$) colour combined with the $J-$/$Ks-$band magnitudes, and following the stellar M/L ratios as a function of optical$-$NIR colours \citep{Bell2003} we estimate the stellar mass of galaxy B to be approximately $3.2-6.8 \times 10^{11} M_\odot$. At this stellar mass, the optical colours are slightly redder ($\sim$0.1 dex) than that of typical red sequence galaxies in low-density environments \citep{Hogg2004}. Both the SFR and stellar mass estimates assume a modified Salpeter initial mass function \citep{Kennicutt1983, Bell2001}. 

In summary, the \mbox{H\,{\sc i}} absorption line detected with ASKAP is associated with an early-type galaxy along the line-of-sight to the radio lobe of PKS\,0409$-$75. The \mbox{H\,{\sc i}} column density is similar to column densities observed in some nearby early-type galaxies, and consistent with \mbox{H\,{\sc i}} emission line studies that show $\sim40\%$ of early-type galaxies show \mbox{H\,{\sc i}} emission \citep{Serra2012}. The optical colours and stellar mass are also similar to that observed in samples of Luminous Red Galaxies (LRGs) at $z\sim0.7$ \citep{Maraston2013, Ching2017}. 

\subsection{Polarisation properties of PKS\,0409$-$75}\label{magneticfield}

Previous narrow-band radio polarization observations of PKS\,0409$-$75 at 3cm and 6cm showed an asymmetry in the fractional polarization and Faraday rotation between the two radio lobes \citep{Morganti1999}. More specifically, the observed Faraday rotation, Rotation Measure ($\rm RM$; the integral of the line-of-sight magnetic field weighted by thermal electron density) of the eastern lobe exceeds that of the western lobe by approximately 300 rad/m$^2$.  While this difference could be produced within or in the vicinity of the lobes, the presence of coherent magnetic fields in the intervening galaxy towards the eastern lobe could also be responsible for the observed Faraday rotation difference. The presence of small-scale turbulent magnetic fields in the intervening galaxy could probably explain the stronger depolarization seen towards the eastern lobe. 

Assuming that the RM difference between the lobes can be fully attributed to the absorber galaxy, and using an ionization fraction of 0.1 \citep{He2013} to convert neutral column density to electron column density, the coherent magnetic field strength, $B$, in the absorber galaxy can be estimated as
\begin{equation}
\frac{B}{14.5\mu {\rm G}} = \left( \frac{\Delta\rm{RM_{rest-frame}}}{804.3~\rm{rad~m^{-2}}} \right) {\left( \frac{T_{\rm{spin}}}{500\rm{K}}\right)}^{-1}.
\end{equation}
A spin temperature T$_{\rm{spin}}$ of 500\,K, rest-frame Faraday rotation difference between the two lobes $\Delta\rm{RM_{rest-frame}}$ of 804.3 rad m$^{-2}$ corresponds to a coherent magnetic field strength of 14.5 $\mu$G in the absorber galaxy. The derived magnetic field strength in the $z = 0.674$ absorber galaxy is broadly consistent with the total magnetic field strengths measured in local spiral galaxies \citep{BeckW2013,Beck2015}, as well as the large-scale coherent disk magnetic field strength measured in a $z=0.439$ galaxy \citep{MaoEA2017}. However, this estimated field strength is higher than that expected in elliptical galaxies \citep{SetaEA2021}. A more in-depth study of the broadband polarimetric properties of PKS\,0409$-$75, including a more detailed estimation of the magnetic fields in the absorber galaxy and its implications on the build-up of galactic magnetic fields over cosmic time will be presented in Seta et al. (in preparation).

\subsection{The detection of \mbox{H\,{\sc i}} absorption against a radio lobe}

Only a handful of \mbox{H\,{\sc i}} absorption lines have been detected against the lobes of radio galaxies, primarily due to selection effects of previous samples that have focused on compact radio sources where the chances of detecting \mbox{H\,{\sc i}} absorption within the host galaxy of the radio source are much greater. The few detections of \mbox{H\,{\sc i}} absorption against a radio lobe have been discovered serendipitously, similar to the case presented here, rather than from a comprehensive survey searching for absorption against extended radio sources. 

A recent study by \citet{Murthy2020} reported \mbox{H\,{\sc i}} absorption against the radio lobe of 3C\,433. In that case, the absorber is a faint disc galaxy with stellar mass of $\sim10^{10}$\,$M_\odot$ blueshifted by only $\sim50$km\,s$^{-1}$ from the systemic velocity of the host galaxy, but has a similar line width of $\sim$80\,km\,s$^{-1}$. Based on this small difference in velocity, combined with some evidence of disturbed morphology in the southern radio lobe led the authors to suggest that the radio lobe could be interacting with the absorbing galaxy. Similar detections of \mbox{H\,{\sc i}} absorption against extended radio sources have been detected in nearby galaxies 3C234 \citep{Pihlstromthesis} and B2 1321$+$31 \citep{Emontsthesis2006}, again where the measured velocity of the \mbox{H\,{\sc i}} absorption signature is very close to the systemic velocity of the radio AGN host galaxy. 

The most distant \mbox{H\,{\sc i}} absorption line against a radio lobe detected prior to this discovery was in the radio galaxy PKS\,1649$-$062 at $z=0.236$ \citep{Curran2011}. In this case the \mbox{H\,{\sc i}} absorption is redshifted 116 km\,s$^{-1}$ compared to the systemic velocity, with the authors suggesting that the absorber is an intervening galaxy gravitationally bound to the system or possibly associated that a large galactic disc with $r\approx200$\,kpc in PKS\,1649$-$062 \citep{Curran2011}. These examples all highlight that the detection of 21-cm absorption lines in radio galaxies can trace a number of different situations, and additional multi-wavelength data, especially in the optical, is often vital in the analysis and interpretation of these systems. 

Unlike these previous detections, in PKS\,0409$-$75 the velocity offset of $\sim3300$\, km\,s$^{-1}$ is too large for the radio lobe to be directly impacting the absorbing galaxy. Instead, the 21-cm absorption line detected towards this powerful radio galaxy is associated with a foreground galaxy in front of the south-eastern radio lobe.  

\subsection{Implications for large absorption line surveys}

Despite being one of the brightest sources in the radio sky, this is the first time \mbox{H\,{\sc i}} absorption has been searched for in this radio source. One reason for this is the fact that as a clearly extended radio source without a strong radio core, PKS\,0409$-$75 would generally not have been included in typical searches for \mbox{H\,{\sc i}} absorption, but a more significant reason is our ability to now search for the 21-cm \mbox{H\,{\sc i}} line in absorption over a wide bandwidth at all redshifts between $0.4<z<1.0$ with the advent of the Australian Square Kilometre Array Pathfinder (ASKAP). This allows us to probe the neutral gas properties of galaxies at these intermediate redshifts for the first time (see e.g. \citealt{Allison2015, Sadler2020, Allison2021}), bridging the gap between \mbox{H\,{\sc i}} studies at lower redshifts (e.g. \citealt{Darling2011}), and QSO Damped Lyman Alpha studies at higher redshifts \citep{Kanekar2014, Neeleman2016, Rao2017}.

The detection of \mbox{H\,{\sc i}} absorption against the lobe of PKS\,0409$-$75 gives some insight into the range of \mbox{H\,{\sc i}} absorption lines that could be detected by upcoming large \mbox{H\,{\sc i}} 21-cm absorption line surveys being carried out with SKA pathfinders such as the First Large Absorption Line Survey in \mbox{H\,{\sc i}} (FLASH; \citealt{flashpaper}) and the MeerKAT Absorption Line Survey (MALS; \citealt{mals}) which will explore intermediate redshifts from $0.4<z<1.5$ in addition to surveys such as the Search for \mbox{H\,{\sc i}} Absorption with AperTIF (SHARP; e.g. \citealt{Adams2019}) and the Widefield ASKAP L-band Legacy All-sky Blind surveY (WALLABY; \citealt{wallaby}) which will detect \mbox{H\,{\sc i}} 21-cm absorption at redshifts $0<z<0.26$. While the sample size presented here is too small to extrapolate the expected detection rate of \mbox{H\,{\sc i}} towards extended radio sources, we note that PKS\,0409-75 is one of five objects in the sample that have clearly extended radio structure with no detectable radio emission from the core. Searching for absorption towards all bright radio sources regardless of underlying source structure over a wide area of sky, such as planned for the upcoming FLASH survey, will provide the first estimate of the typical detection rate towards extended radio sources. This study also highlights the need for deep multi-wavelength follow-up observations to understand the properties of \mbox{H\,{\sc i}} absorbers that will be detected in these large surveys. 

\section{Conclusions}

We have carried out a search for 21-cm \mbox{H\,{\sc i}} absorption against 16 bright southern radio sources with $0.2<z<0.7$ using ASKAP-BETA, the commissioning array of the Australian Square Kilometre Array Pathfinder. This study revealed two detections of \mbox{H\,{\sc i}} absorption; one in the compact steep spectrum source PKS\,0023$-$26 at $z=0.322$ and another towards the radio galaxy PKS\,0409$-$75. The former shows similar line properties to that reported by \citet{Vermeulen2003}, with no significant evidence for variability in the \mbox{H\,{\sc i}} line. The absorption towards PKS\,0409$-$75 at $z=0.674$ has not been detected previously and is blueshifted from the systemic velocity of the host galaxy by $\sim 3300$\,km\,s$^{-1}$. Follow-up optical imaging and spectroscopy with GMOS on Gemini-South confirmed that the \mbox{H\,{\sc i}} absorption is associated with a foreground galaxy in front of the eastern radio lobe. Using the optical and radio data combined we estimate this galaxy to have a stellar mass of $3.2 - 6.8 \times 10^{11}M_\odot$ and an \mbox{H\,{\sc i}} column density of $2.16\times10^{21}$\,cm$^{-2}$. Using archival polarisation measurements of PKS\,0409$-$75 from \citet{Morganti1999}, we provide an estimate of the magnetic field in the foreground galaxy of $\sim 14.5\mu$G, giving the first estimate of the magnetic field of a normal galaxy at $z\sim0.7$. 

Although only a small sample was searched for \mbox{H\,{\sc i}} absorption in this work, it provides a pilot study of what could be detected by future large absorption line surveys such as the First Large Absorption Line Survey in \mbox{H\,{\sc i}} \citep{flashpaper}, currently being carried out with ASKAP. Having a sufficiently bright background radio source was the primary selection criteria used to define this sample, with no further constraints on the source size or optical properties (with the exception that the \mbox{H\,{\sc i}} absorption line would be in the frequency range observable with ASKAP). The result of this broader selection criteria led to the detection of absorption towards the lobe of PKS\,0409$-$75 as its extended radio structure and large velocity offset from the host galaxy means that it would likely have been excluded from many previous searches for \mbox{H\,{\sc i}} absorption.

\section*{Acknowledgements}

We thank Jesse van der Sande for useful discussions on the optical data analysis presented here and the anonymous referee for helpful suggestions that improved the quality of the paper. Optical observations for this paper were carried out under the Gemini programme GS-2017B-Q-63 and we thank the telescope operators and staff for carrying out these observations. We also thank CSIRO operations and engineering staff who assisted in obtaining the ASKAP-BETA observations. Parts of this research were conducted by the Australian Research Council Centre of Excellence for All-sky Astrophysics in 3D (ASTRO 3D) through project number CE170100013.

The Australian SKA Pathfinder is part of the Australia Telescope National Facility which is managed by CSIRO. Operation of ASKAP is funded by the Australian Government with support from the National Collaborative Research Infrastructure Strategy. ASKAP uses the resources of the Pawsey Supercomputing Centre. Establishment of ASKAP, the Murchison Radio-astronomy Observatory and the Pawsey Supercomputing Centre are initiatives of the Australian Government, with support from the Government of Western Australia and the Science and Industry Endowment Fund. We acknowledge the Wajarri Yamatji people as the traditional owners of the Observatory site. 

This research made use of APLpy, an open-source plotting package for Python \citep{aplpy2012,aplpy2019} and  Astropy,\footnote{http://www.astropy.org} a community-developed core Python package for Astronomy \citep{astropy2013, astropy2018}.

\section*{Data Availability}

All data underlying this article will be shared on reasonable request to the corresponding author. The optical data are publicly available for download through the Gemini archive under project code GS-2017B-Q-63 at https://archive.gemini.edu/searchform. 

%%%%%%%%%%%%%%%%%%%%%%%%%%%%%%%%%%%%%%%%%%%%%%%%%%



\bibliographystyle{mnras}
\bibliography{2jy} % if your bibtex file is called example.bib

%%%%%%%%%%%%%%%%%%%%%%%%%%%%%%%%%%%%%%%%%%%%%%%%%%

%%%%%%%%%%%%%%%%% APPENDICES %%%%%%%%%%%%%%%%%%%%%

\appendix

\section{ASKAP-BETA Observations}

ASKAP-BETA observations of 16 sources selected from the 2-Jy sample of southern radio galaxies were observed between November 2014 and February 2016. Table \ref{obstab} details the observation data, schedule block ID number (SBID) and median spectral noise across the full bandwidth for each observation of this sample.

\begin{table*}
\caption{ASKAP-BETA observations of sources selected from the 2-Jy sample. Columns are (1) Source name, (2) date of observation, (3) ASKAP-BETA Scheduling Block ID (SBID), (4) duration of observations, (5) median spectral noise across the full band, (6) median fractional absorption noise (i.e. $\sigma$/$S_{\rm cont}$). \label{obstab}}
\begin{tabular}{llcccc}
\hline
Source & Date & SBID & t & $\sigma$ & $\sigma$/$S_{\rm cont}$ \\
& & & (hrs) & mJy/bm/ch & (\%) \\
\hline
%{\it A-priority sample:} & & & & & & & \\
\multicolumn{4}{l}{\it Band 1 observations: (711.5 - 1015.5 MHz)} \\ 
PKS\,0105$-$16 & 2015-sep-11 & 2538 & 4.5 & 43.3 & 0.68 \\
 & 2015-sep-11 & 2541 & 4.5 & 58.0 & 0.86 \\
%PKS\,0105$-$16 & 2015-sep-11 & 2541 & 4.5 & 0.04 & R \\
 & 2015-dec-10 & 3302 & 2.5 & 33.6 & 1.02 \\
%PKS\,0117$-$15 & 2015-jun-25 & 1994 & 6.5 & find & \\
& {\bf Total} & & {\bf 7.0} & & {\bf 0.47}  \\
PKS\,0117$-$15 & 2015-jun-25 & 1998 & 6.5 & 35.5 & 0.53  \\
 & 2015-sep-12 & 2546 & 3.0 & 57.5 & 0.72  \\
 & 2015-dec-10 & 3302 & 2.5 & 36.3 & 0.47  \\
& {\bf Total} & & {\bf 12.0} & & {\bf 0.31}  \\
PKS\,0235$-$19 & 2014-sep-15 & 636 & 3.0 & 28.4 & 0.40  \\
 & 2015-dec-10 & 3302 & 3.0 & 38.9 & 0.52  \\
& {\bf Total} & & {\bf 6.0} & & {\bf 0.31} \\
PKS\,0252$-$71 & 2014-nov-11 & 1070 & 3.5 & 22.6 & 0.23 \\
%PKS0252$-$71 & 2015-aug-25 & 2450 & 3.0 & - & S \\
& 2015-sep-12 & 2546 & 3.0 & 57.3 & 0.61  \\
 & 2015-oct-01 & 2675 & 5.0 & 26.0 &  0.28 \\
 & 2015-oct-02 & 2680 & 4.0 & 25.9 &  0.28 \\
 & 2015-dec-12 & 3317 & 3.0 & 28.7 & 0.31 \\
 & 2015-dec-13 & 3321 & 6.0 & 20.2 & 0.22  \\
& {\bf Total} & & {\bf 24.5} & & {\bf 0.11}  \\
PKS\,0409$-$75 & 2014-nov-20 & 1143 & 3.5 & 45.6 & 0.10  \\
%PKS\,0409$-$75 & 2015-mar-18 & 1549 & 5.0 & find & \\
%PKS\,0409$-$75 & 2015-mar-19 & 1551 & 6.0 & find & \\
 & 2015-may-05 & 1753 & 4.0 & 20.7 & 0.11  \\
 & 2015-aug-28 & 2483 & 3.0 & 29.2 & 0.15  \\
 & 2015-sep-19 & 2591 & 5.0 & 24.6 & 0.11  \\
 & 2015-sep-30 & 2668 & 5.0 & 22.2 & 0.11  \\
 & 2015-oct-04 & 2687 & 6.0 & 21.6 & 0.10  \\
%PKS\,0409$-$75 & 2015-oct-20 & 2791 & 8.0 & 0.005 & S \\
 & 2016-feb-14 & 3577 & 7.0 & 23.5 & 0.13  \\
& {\bf Total} & & {\bf 41.5} & & {\bf 0.05}  \\
PKS\,1136$-$13 & 2014-nov-22 & 1147 & 2.0 & 34.0 & 0.50  \\
 & 2014-nov-23 & 1150 & 2.0 & 32.5 & 0.48  \\
 & 2015-dec-18 & 3394 & 3.0 & 29.8 & 0.49  \\ 
 & 2016-jan-28 & 3469 & 4.0 & 51.2 & 0.80  \\
 & 2016-jan-29 & 3470 & 4.0 & 47.8 & 0.81  \\
 & 2016-feb-03 & 3500 & 3.0 & 55.8 & 0.87  \\ 
& {\bf Total} & & {\bf 18.0} & & {\bf 0.25}  \\
PKS\,1306$-$09 & 2014-nov-20 & 1143 & 3.5 & 54.9 & 0.60 \\
%PKS\,1306$-$09 & 2015-jun-28 & 2012 & 4.0 & find &\\
 & 2015-jul-11 & 2093 & 3.0 & 50.7 & 0.61 \\
 & 2015-oct-03 & 2681 & 4.0 & 26.8 & 0.49 \\
 & 2015-dec-10 & 3303 & 3.5 & 33.3 & 0.52 \\
 & 2015-dec-19 & 3399 & 3.5 & 30.4 & 0.51  \\
 & 2016-jan-28 & 3469 & 7.0 & 51.0 & 0.87 \\
 & 2016-jan-29 & 3470 & 5.0 & 42.7 & 0.79  \\
 & 2016-jan-30 & 3475 & 6.0 & 42.5 & 0.78  \\
& {\bf Total} & & {\bf 35.5} & & {\bf 0.21} \\
PKS\,1547$-$79 & 2014-nov-14 & 1096 & 3.0 & 23.6 & 0.51  \\
%PKS\,1547$-$79 & 2015-mar-11 & 1528 & 2.5 & ? & find \\
 & 2016-jan-28 & 3469 & 5.0 & 42.1 & 0.88  \\
& {\bf Total}& & {\bf 10.5} & & {\bf 0.44} \\
PKS\,1602$+$01 & 2015-jul-11 & 2093 & 4 & 43.8 & 0.51  \\
 & 2015-dec-10 & 3308 & 4 & 29.0 & 0.43  \\
 & 2015-dec-19 & 3399 & 3.5 & 30.0 & 0.42  \\
& {\bf Total} & & {\bf 11.5} & & {\bf 0.26}  \\
PKS\,1938$-$15 & 2015-jul-14 & 2127 & 3.0 & 32.9 & 0.31  \\
 & 2015-sep-12 & 2546 & 3.0 & 43.7 & 0.44  \\
& {\bf Total} & & {\bf 6.0} & & {\bf 0.41} \\
PKS\,2135$-$20 & 2016-jan-28 & 3468 & 2.5 & 67.1 & 1.26  \\
 & 2016-jan-29 & 3472 & 2.0 & 68.7 & 1.31  \\
& 2016-jan-30 & 3480 & 3.0 & 57.4 & 1.11  \\
& {\bf Total} & & {\bf 7.5} & & {\bf 0.69}  \\
\multicolumn{4}{l}{\it Band 2 observations: (967.5--1271.5 MHz)} \\ 
PKS\,0023$-$26 & 2015-nov-23 & 3184 & {\bf 3.0} & {\bf 50.1} & {\bf 0.50}  \\
PKS\,0035$-$02 & 2015-nov-24 & 3190 & {\bf 4.0} & {\bf 94.9} & {\bf 1.68}  \\
PKS\,0039$-$44 & 2015-nov-23 & 3185 & {\bf 3.0} & {\bf 84.8} & {\bf 1.68} \\
PKS\,1151$-$34 & 2015-nov-23 & 3182 & {\bf 3.0} & {\bf 66.4} & {\bf 0.97}  \\
%PKS\,1151$-$34 & 2015-nov-23 & 3185 & 3.0 & band 2, - & R \\
PKS\,2135$-$14 & 2015-nov-23 & 3184 & {\bf 3.0} & {\bf 52.1} & {\bf 3.10}  \\
\\
\hline
\end{tabular}
\end{table*}

\section{ASKAP-BETA spectra }

\begin{figure*}
\centering{\includegraphics[width=0.9\linewidth]{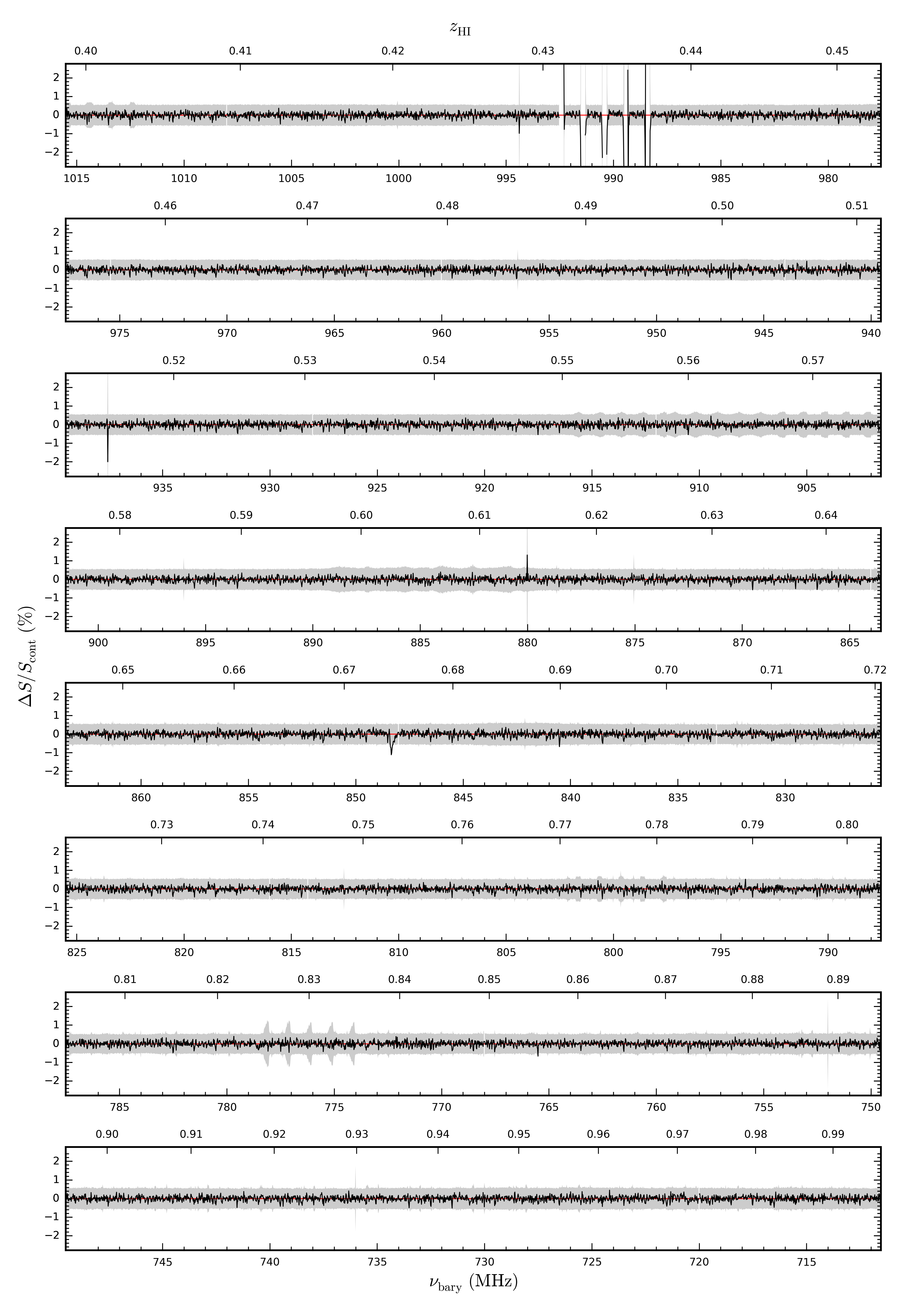}}
\caption{Full spectrum of a 5 hr observation of PKS\,0409$-$75 observed on September 30, 2015. The grey band signals the 5$\sigma$ noise level calculated from the rms of the image per channel. \mbox{H\,{\sc i}} absorption is clearly detected at a frequency of 848\,MHz. Some single channel noise peaks are seen occasionally (e.g. at 937.5\,MHz) as well as artefacts due to correlator errors such as that seen around 990\,MHz and 776\,MHz. \label{fullspectrum}}
\end{figure*}

ASKAP-BETA provided a full bandwidth of 304\,MHz covering a redshift range of $0.4<z<1$ for objects observed at a central frequency of 863.5\,MHz.  Figure \ref{fullspectrum} shows an example spectrum obtained from one observation of PKS\,0409-75 demonstrating the full bandwidth covered for all observations and the lack of radio frequency interference at these frequencies. The top axis shows the redshift of the 21-cm \mbox{H\,{\sc i}} line and the bottom axis shows the frequency shifted into the barycentric reference frame ($\nu_{\rm bary}$).

% Don't change these lines
\bsp	% typesetting comment
\label{lastpage}
\end{document}